\documentclass[superscriptaddress,aps,10pt,showkeys,notitlepage,twocolumn,prb,reprint,longbibliography,floatfix]{revtex4-2}
\usepackage{natbib}
\usepackage{times}
\usepackage[colorlinks,bookmarksopen,bookmarksnumbered,linkcolor=cyan,citecolor=cyan,urlcolor=blue,breaklinks=true]{hyperref}
\usepackage{bm}
\usepackage[english]{babel}
\usepackage{amssymb}
\usepackage{amsmath}
\usepackage{amsthm}
\usepackage{epsfig}
\usepackage{array,multirow,makecell}
\usepackage{url}
\usepackage{xcolor}
\usepackage{setspace}
\usepackage{anyfontsize}
\usepackage{hyperref}

\newcommand{\av}[1]{\left\langle #1\right\rangle}  
\newcommand{\bigO}[1]{{\rm O}\left(#1\right)}      
\newcommand{\dd}{\mathrm{d}} 
\newcommand{\ee}{\mathrm{e}} 
\newcommand{\ii}{\mathrm{i}} 
\newcommand{\Pf}{\mathop{\mathrm{Pf}}}   
\newcommand{\pv}{\mathop{\mathrm{p.v.}}} 
\newcommand{\sign}{\mathop{\text{sign}}} 
\newcommand{\cL}{c_{\text{L}}}           
\newcommand{\cR}{c_{\text{R}}}           
\newcommand{\cS}{c_{\text{S}}}           
\newcommand{\etae}{\eta_{\text{e}}}      
\newcommand{\siga}{\sigma_{\text{a}}}    
\newcommand{\sigc}{\sigma_{\text{c}}}    
\newcommand{\sigth}{\sigma_{\text{th}}}  
\begin{document}
\title{Shock-driven nucleation and self-organization of dislocations in the dynamical Peierls model}
\author{Y.-P. Pellegrini}
\email{yves-patrick.pellegrini@cea.fr} 
\affiliation{CEA, DAM, DIF, F-91297 Arpajon, France.}
\affiliation{Universit{\'e} Paris-Saclay, LMCE, 91680 Bruy{\`e}res-le-Ch{\^a}tel, France}
\author{M. Josien}
\email{marc.josien@cea.fr} 
\affiliation{CEA Cadarache, F-13108 St-Paul-lez-Durance, France.}

\begin{abstract}
Dynamic nucleation of dislocations caused by a stress front ('shock') of amplitude $\siga$ moving with speed $V$ is investigated by solving numerically the dynamic Peierls equation with an efficient method. Speed $V$ and amplitude $\siga$ are considered as independent variables, with $V$ possibly exceeding the longitudinal wavespeed $\cL$. Various reactions between dislocations take place such as scattering, dislocation-pair nucleation, annihilation, and crossing. Pairs of edge dislocation are always nucleated with speed $v\gtrsim\cL$ (and likewise for screws with $\cL$ replaced by $\cS$, the shear wavespeed). The plastic wave exhibits self-organization, forming distinct `bulk' and `front' zones. Nucleations occur either within the bulk or at the zone interface, depending on the value of $V$. The front zone accumulates dislocations that are expelled from the bulk or from the interface. In each zone, dislocation speeds and densities are measured as functions of simulation parameters. The densities exhibit a scaling behavior with stress, given by $((\siga/\sigth)^2-1)^\beta$, where $\sigth$ represents the nucleation threshold and $0<\beta<1$.
\medskip

\emph{To appear in Physical Review B}
\end{abstract}
\date{\today}
\maketitle
\section{Introduction}
Shock loading of metals \cite{MOGI83,KANE04,*FORB12} involves a high rate of dislocation \cite{ANDE17} generation \cite{ARMS10b,AUST12}. Realistically accounting for it remains a challenge for dislocation-based multiscale models of shock-loaded solids aimed at very high strain rates \cite{AUST12,BART13,KOSI19,*KOSI21a,YAOY21a,*YAOP22,YELI23}, in spite of a vast amount of data from atomistic simulations \cite{HOLI98,GERM00,*GERM04a,TANG03,MEYE09,ZHAK11a,SICH16,BISH19}. Shock-induced plasticity has also been explored using two-dimensional, non-supersonic, elastodynamic dislocation simulations with retarded interactions and separate nucleation criterion and mobility law \cite{GURR15a,*GURR15c,*GURR15e}. At high strain rates, line multiplication from traditional sources \cite{SHEH06,*BEIG08} via the Orowan mechanism is no more the dominant generation mechanism \cite{SCHI95,*SCHI02}, and nucleation must be considered instead \cite{GURR15a,*GURR15c,*GURR15e}. Whereas standard thermally-activated homogeneous nucleation \cite{SHEH16} is usually invoked below the nucleation stress threshold, no consensus exists for larger driving stresses. However newly-nucleated dislocations can trigger further nucleation events \cite{LOMD86,BAGG23a,*BAGG23b}. 

Most dislocation-based models rely on the assumption that the plastic strain rate is proportional to the density and speed of mobile dislocations by Orowan's equation \cite{OROW40,TAYL65}. Thus, speed is also a much debated issue. In uniaxial compression, the Hugoniot elastic limit \cite{KANE04,DAVI08} where plasticity comes into play is first attained for shear on planes inclined $45^\circ$ from the compression axis. Dislocations accompanying a compression wave of (longitudinal) speed $\cL$ should then move with supersonic speed $\smash{v=\sqrt{2}\,\cL}$ \cite{MEYE78,WEER81b}. Early models \cite{SMIT58,*HORN62} postulated layers of intersonic or supersonic edge dislocations \cite{WEER67b,*WEER81b,*WEER88} at the shock front to accomodate shock-induced compression, with residual dislocation density in the bulk. To dispense with supersonic dislocations, alternative proposals \cite{MEYE78,MEYE94} considered dislocation loops homogeneously and repeatedly nucleated behind the shock front by the deviatoric stress set up by uniaxial strain, moving thereafter only short distances at subsonic speeds. Scarce information is available concerning supersonic dislocation speeds in shocks, and in absence of direct evidence the experimental status of supersonic dislocations in metals remains unsettled \cite{MEYE94}. Still, atomistic simulations have revealed that intersonic or supersonic dislocations are quite possibly involved \cite{GUMB99b,*JIN08,TSUZ08,HAHN16}, although current dislocation-based density evolution models disregard this possibility.

Leaving aside thermal effets \cite{DAVI08,RYUK11,GURR17b}, which are presumably irrelevant \cite{WEER81b} at high stress except for changes in elastic constants, wavespeeds, and phonon drag, the present work investigates nucleation of perfect dislocations in a one-dimensional (1D) idealized model for straight dislocations on a single slip plane, with driving stress exceeding the nucleation threshold, and variable `shock' speed. To this aim we use the (elasto-) Dynamic Peierls Equation (DPE) \cite{PELL10b,PELL14}, which we solve numerically \cite{JOSI18b}. This equation for the plastic slip rests on minimal assumptions within the small-strain formulation of continuum mechanics, assuming fixed elastic constants. Its properties with regard to dislocation nucleation and collective behavior are unexplored so far. By construction, it accounts for stress-wave emission and scattering \cite{MAUR04b} by/on dislocations in elastodynamic interaction. Dislocations are free to glide (i.e., all mobile) with no obstacles, nor Peierls stress. Supersonic motion is allowed. Once nucleated as pairs of opposite signs \cite{MEYE78}, they interact via retarded elastodynamic interactions. Dislocation trajectories as well as quantities usually considered in constitutive models, such as the plastic strain rate $\dot{\varepsilon}^{\rm p}$, dislocation density $\rho$, speed $v$ and stress, are monitored. Our approach is motivated by simplicity and a direct connection between this continuum description and Orowan's equation. More accurate analogous lattice-based dislocation models \cite{GURR19c} might be less suitable to this purpose. 

The DPE is reviewed and cast in Sec.\ \ref{sec:dpe} in a form suitable to numerical solution (see Appendix \ref{sec:methsol} for the method). Sec.\ \ref{sec:shocknucl} is devoted to simulations. After the setup is explained, the plastic structure arising from dislocation trajectories is discussed, with mean dislocation speeds and speed probability densities measured. Nucleation is addressed next, and mean dislocation densities are computed over a wide range of `shock' speeds and driving stresses. A scaling in stress of the density is proposed on the basis of a supersonic-train solution derived in Appendix \ref{sec:sda}.  Concluding remarks close the paper (Sec.\ \ref{sec:concl}). Appendices are devoted to the numerical method. Our numerical results privilege the edge dislocation, as its behavior is more involved than the screw's. Additional results for screws are reported in the Supplemental Material \cite{SUPP}.

\section{Dynamical Peierls Equation}
\label{sec:dpe}
\subsection{Overview and former results}
\label{sec:dpe_overview}
\begin{figure}[!ht]
	\centering
	\includegraphics[width=6cm]{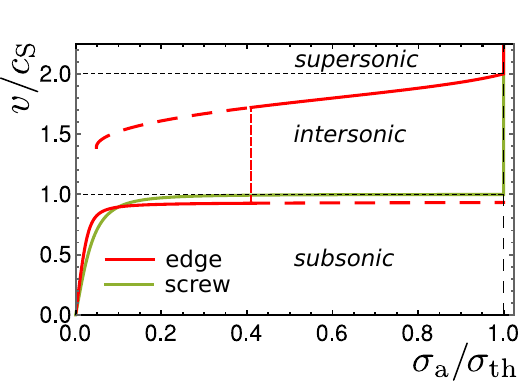}
	\caption{\label{fig:fig1}(Color online) Single-dislocation steady-state stable bran\-ches of the mobility law for screw and edge dislocation in the PND model, for $\cL/\cS=2$ and $\alpha=0.05$ (see below for a definition of this quantity). Edge case: the solid-line part depicts states accessible under single-step loading; the dashed part depicts states accessible under double-step loading \cite{PELL14}. The jump between subsonic and intersonic branches occurs at $\siga=\sigc$ (see text).}
\end{figure}

The DPE extends the classical Peierls model of a dislocation \cite{PEIE40,*NABA47,ANDE17} to elastodynamics \cite{ESHE53b,MARK08,LAZA16a,CUIP19}. It reduces to the Weertman equation \cite{WEER69a,*WEER69b,ROSA01,JOSI18a} for steady motion under uniform applied stess $\siga$. In Ref.\ \onlinecite{PELL14} -- hereafter referred to as (I), and for a sinusoidal (Frenkel) pull-back force, the elastodynamic solution of the DPE was approached via a collective-variable approximation (CVA), using a single-dislocation ansatz depending on two collective variables -- dislocation position and width, for which evolution equations were obtained. The CVA is such that its stable steady-velocity states at large times match those of the DPE for one dislocation, which are given by the single-dislocation solution of the Weertman equation. The resulting mobility law \cite{ROSA01,PELL14} is recalled in Fig.\ \ref{fig:fig1}, where $\sigth=\mu b/(2\pi d)\sim \mu/10$, is the theoretical shear stress (lattice shear strength) \cite{OROW40,OGAT04,ANDE17} defined in terms of the shear modulus $\mu$, the Burgers vector $b$ and the interplane distance $d$ \cite{ANDE17}. Hereafter, $\cS$ and $\cL$ denote the shear and longitudinal wave speeds of the medium. The screw dislocation has only one subsonic branch that saturates at $v=\cS$. The edge dislocation has two stable branches: one subsonic ($|v|<\cS$), the other one intersonic ($\cS<|v|<\cL$). The subsonic branch saturates at the Rayleigh velocity $v=\cR\simeq 0.93\,\cS$. No single-dislocation mobility law exists for $\siga>\sigth$ in this model, and the speed is undetermined if $\siga=\sigth$. The inclusion of a gradient term provides a means to overcome this limitation \cite{ROSA01}. However, our study demonstrates that under high-stress `shock' conditions the DPE yields multi-dislocation supersonic solutions \emph{without needing any such modification}.

Atomistic calculations for an edge dislocation \cite{GUMB99b,*JIN08} suggested a dependence of branch selection on the way the external loading is applied. This was further investigated in (I), leading to the following predictions: a) when $\siga$ is suddenly applied and thereafter maintained (single-step loading), a critical stress $\sigc$ exists that separates regimes with subsonic or intersonic terminal dislocation speeds. This transition is represented at $\sigc\simeq 0.41\,\sigth$ in Fig.\ \ref{fig:fig1} Thus the model accounts for the jumps from one velocity branch of the mobility law to the other as stress increases; b) the terminal speed is dynamically selected and depends on the history of loading. Two-step loading, where a first stress level is imposed and subsequently modified after some time, grants access to asymptotic states forbidden under single-step loading \cite{PELL14}. In anisotropic materials more branches come into play \cite{MARI06,OLMS05,TSUZ08,DAPH14,OREN17}.
  
\subsection{The equation}
\label{sec:theeq}
The DPE for the plastic slip function $\eta(x,t)$ of a system of straight dislocations moving along the $x$-axis reads
\begin{align}
\label{eq:dpe}
-\frac{\mu}{\pi}&\int \dd x'\int_{-\infty}^{t}\dd t' K_a(x-x',t-t')\frac{\partial\eta}{\partial x}(x',t')\\
&{}-\kappa_a^\alpha\frac{\mu}{2\cS}\frac{\partial\eta}{\partial t}(x,t)+\siga(x,t)=f\bigl(\eta(x,t)+\etae(x,t)\bigr).\nonumber
\end{align}
The left-hand side of this equation represents the \emph{local stress} $\sigma(x,t)$ on the slip plane. The quantity $\siga(x,t)$ is an externally applied stress, which depends \emph{on both position and time}. The local stress is balanced by the pull-back force $f(\eta)$, which derives from the generalized stacking-fault $\gamma$-potential of the lattice \cite{CHRI70a}.
We use here the Frenkel force $f(\eta)=\sigth\sin(2\pi\eta/b)$, which implies perfect dislocations, but our numerical method applies as well to any other force. This force favors locking of $\eta$ on integer multiples of $b$. Dislocations lie where a transition between two successive multiples take place. The theoretical shear stress $\sigth$ is its maximal value, which identifies with the nucleation stress \cite{WEER81b,RICE92b,*XUAR95,HOLI98} (see Sec.\ \ref{sec:nucleation} below). 
Equation \eqref{eq:dpe} implies that the local stress in the system is \emph{always limited by $\sigth$, that is, $|\sigma(x,t)|\leq\sigth$, whatever $\siga$}.
In \eqref{eq:dpe} the function $\etae$ represents the elastic part of the slip, i.e., the dislocation-free solution of the equation $f=\siga$ if $|\siga|\leq \sigth$. It increases with $\siga$ up to its saturation value $b/4$ for $\siga=\sigth$, and by convention remains fixed to that value if $\siga>\sigth$ \cite{PELL14}. Kernel $K_a(x,t)$ accounts for retarded elastodynamic self-interactions on the slip plane of the field $\eta(x,t)$. It depends on the dislocation character $a=\text{s}$ (screw), $\text{g}$ (glide edge), $\text{c}$ (climb edge). The local instantaneous term with $\partial\eta/\partial t$ is proportional to the dimensionless constant \footnote{\protect In Refs.\ \onlinecite{PELL14,JOSI18a}, this coefficient was defined as $\kappa_a^\alpha=\kappa_a+\alpha$. The present modification only (formally) impacts the \emph{climb} kernel. It rationalizes the pole structure of the kernels in the plane $(\alpha,\gamma)$, where the value $\alpha=1$ is special; see Supplemental Material \cite{SUPP}.}
\begin{align}
\label{eq:kappaaa}
\kappa_a^\alpha=\kappa_a(1+\alpha).
\end{align}
Whereas $\kappa_a$ embodies radiative losses associated with $K_a$, the contribution $\alpha\kappa_a$ with $\alpha>0$ accounts phenomenologically for phonon drag. The inclusion of the phonon-drag term ensures a well-defined mobility law in the subsonic range depicted in Fig.\ 1, while also providing the required linear mobility behavior at low speeds \cite{WEER69b,ROSA01,PELL14}. Expressions of $K_a(x,t)$ and $\kappa_a$ are reported in Appendix \ref{sec:kernels}. It is worth noting that the elastic waves propagated by $K_{\rm a}$ undergo no damping. Indeed, bulk elasto-viscosity, or anelasticity, is insignificant in pure pristine metal crystals \cite{ESHE61,*NOWI72,*MEYE77}. Additionally, the potential inclusion of inertia in the pull-back force \cite{ESHE53b}, while unexplored at present and lacking documentation, is intentionally disregarded in order to maintain tractability of equations in the steady state. The emphasis in the current work is placed on simplicity to enable improved control and facilitate interpretations based on existing references \cite{ROSA01,PELL14}. 
 
Evolution starts at $t=0$. Prior to this instant, $\siga(x,t)=0$, and the system can, e.g., be assumed to contain one dislocation at rest, of shape $\eta_0$. The initial state $\eta_0(x)\equiv \eta(x,0)$ is a solution of the Peierls-Nabarro (PN) equation
\begin{align}
\label{eq:pn}
-\frac{\mu\mathcal{K}_a}{\pi}\pv\int \frac{\dd x'}{x-x'}\frac{\partial\eta_0}{\partial x}(x')&=f\bigl(\eta_0(x)\bigr),
\end{align}
where $\pv$ stands for Cauchy's principal value, $\mathcal{K}_\text{s}=1/2$ or $\mathcal{K}_\text{g,c}=1/[2(1-\nu)]$, and $\nu$ is Poisson's ratio. Indeed the DPE equation reduces to the latter in statics \cite{PELL10b}.

The dynamic solution is defined relatively to this state as
\begin{align}
\label{eq:etasplit}
\eta(x,t)=\eta_0(x)+\delta\eta(x,t), 
\end{align}
where $\delta\eta(x,t)\equiv 0$ for $t\leq 0$. For $t>0$ the equation to be solved for $\delta\eta(x,t)$ is, with same dependency on $(x,t)$ as above  
\begin{subequations}
\begin{align}
\label{eq:dpe2}
-\frac{\mu}{\pi}&\int \dd x'\int_{-\infty}^{t}\dd t' K_a\frac{\partial\delta\eta}{\partial x}
-\kappa_a^\alpha\frac{\mu}{2\cS}\frac{\partial\delta\eta}{\partial t}=F,
\end{align}
where $F(x,t)$ is
\begin{align}
\label{eq:Fdef}
F\equiv f(\eta_0+\delta\eta+\eta_{\rm e})-f(\eta_0)-\siga.
\end{align}
\end{subequations}
In absence of initial dislocation, one simply takes $\eta_0(x)\equiv 0$.

Throughout the paper, the longitudinal wavespeed is $\cL=2\,\cS$ (which corresponds to a Poisson ratio $\nu=1/3$) and the drag coefficient is $\alpha=0.01$ \cite{PELL14}. Results will be given in dimensionless form, with time in units of $\tau_0=d/\cS$, positions in units of $d$, speeds in units of $\cS$, and stresses in units of $\sigth$ so that the values of $\mu$, $b$, and $d$ need not be precised. 

The numerical method of solution is sketched in Appendix \ref{sec:methsol}. Prior to going further, the numerical solution of the DPE for one pre-existing single dislocation has been cross-checked against the collective-variable approximation with good overall agreement. In particular, predictions of the CVA regarding the subsonic-intersonic transition for an edge dislocation \cite{PELL14} are fully confirmed (see Supplemental Material \cite{SUPP}). 

\subsection{Local form of Orowan's equation}
The `microscopic' dislocation density $\rho$ and plastic strain rate $\dot{\varepsilon}^{\rm p}$ are 
\begin{align}
\label{eq:parder}
\rho&=-\frac{1}{d b}\frac{\partial\eta}{\partial x}, \qquad \dot{\varepsilon}^{\rm p}=\frac{1}{d}\frac{\dd\eta}{\dd t},
\end{align} 
where $d$ is the interplane distance \cite{ANDE17}. If we neglect core-shape changes and let $\eta(x,t)=\eta(x-\xi(t))$ represent a dislocation train moving  with velocity $v=\dot{\xi}$, a local instance of Orowan's equation \cite{OROW40} is retrieved from \eqref{eq:parder} as
\begin{align}
\label{eq:locorowan}
\dot{\varepsilon}^{\rm p}&=b \rho v.
\end{align}
However, accounting for dynamical core-shape changes via an intrinsic time-dependence of $\eta$ yields the more general relationship 
\begin{align}
\label{eq:locorowangen}
\dot{\varepsilon}^{\rm p}&=b \rho v+b\rho \frac{\partial\eta}{\partial t}.
\end{align}
Upon ensemble and time averaging, this expression should produce a correction to \eqref{eq:locorowan} of a type different from the one, of density-rate origin, proposed by Armstrong and Zerilli \cite{ARMS10b,AUST12} to account for nucleation. Indeed, Orowan's equation \eqref{eq:locorowan} was derived assuming a constant dislocation density \cite{OROW40}. Nevertheless, this equation can always (formally) be enforced at the macroscopic level by \emph{defining} the mean speed as
\begin{align}
\label{eq:globorowan}
\overline{v}&:=\overline{\dot{\varepsilon}}^{\rm p}/(b\overline{\rho}),
\end{align}  
where the overbar encompasses all relevant averages \footnote{A particular instance of the latter definition is implicit in the rectangular loop model \cite{JOHN59}.}. With Eqs.\ \eqref{eq:locorowangen} and \eqref{eq:globorowan}, the mean speed may turn out very different from that resulting from the mobility law of individual dislocations. 

\section{`Shock' simulations}
\label{sec:shocknucl}
\subsection{Set-up}
\begin{figure}[!ht]
	\centering
	\includegraphics[width=5.5cm]{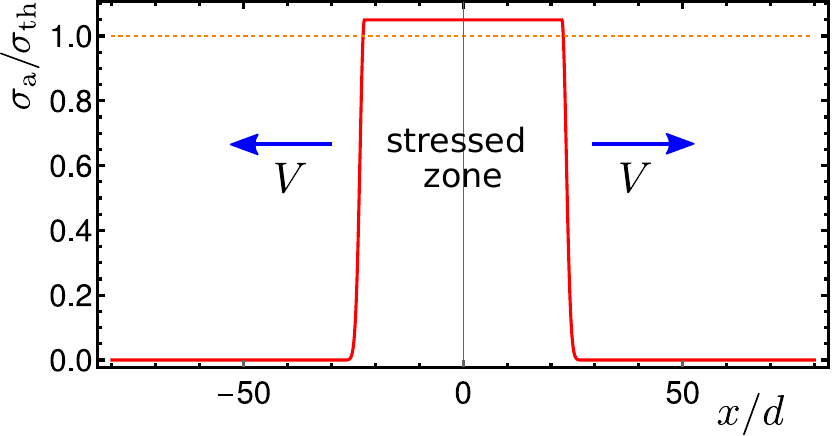}
	\caption{\label{fig:fig2} (Color online) Time-dependent applied stress profile $\siga(x,t)$ expanding with speed $V$, for  $\siga=1.05$.}
\end{figure}
Highly-simplified `shock'-like conditions are simulated by uniformly loading with stress $\smash{\siga>\sigth}$ a region expanding from the center of the system with expansion speed $V$ (hereafter referred to as `shock speed'); see Fig.\ \ref{fig:fig2}(a). Hereafter $\siga$ stands as the plateau value of the function $\siga(x,t)$. The stress varies spatially from $0$ to $\siga$ over a small width constant over time. Thus, for simplicity, our `shocks' are merely moving stress steps. For investigation purposes, we assume $\smash{\siga}$ and $V$ to be independent parameters. In shock physics, however, these parameters are connected via the pressure/shock speed relation determined by the Rayleigh line on the Hugoniot \cite{KANE04,*FORB12,DAVI08}. 
The parametric domain explored is $\smash{1.05\leq \siga/\sigth\leq 2.0}$ (11 values) and $0.1\leq V/\cS \leq 3.0$ (31 values), using $\cL=2\cS$. If $V$ is moderate with respect to wavespeeds, the loading operates over an expanding region as does a nanoindentation process in pressure \cite{LORE03,MASO06,BEIG08,MRID19}; if instead $V>\cL$ for edges (resp.\ $\cS$ for screws), it mimicks a shock of speed exceeding the upper wavespeed relevant to the dislocation character considered. In shock physics the regime $V>\cL$ where the shock overruns elastic waves is often referred to as \emph{overdriven} \cite{WALL81a,PRES03,ZHAK11a}. We emphasize that physical shocks possess inherent nonlinear characteristics that are beyond the scope of our `shock' model, as further discussed in the concluding Sec. \ref{sec:concl}. For the time being, we will omit the quotation marks around the term `shock', although it is crucial to remain aware of the simplifications made with regard to shock physics.  

Simulations have been carried out on systems of size $L=(160\pi)d$ for screw dislocations, and $L=(320\pi)d$ for edge dislocations, discretized into 4096 spatial points, with periodic boundary conditions (PBCs), and inner time step $\smash{\delta t= 3.\,10^{-2}\tau_0}$ further subdivided by the algorithm when necessary. Derivatives $\partial\eta/\partial x$ and $\partial\eta/\partial t$ are estimated by finite differences using the inner time step. For each run, the various components of the DPE (self-stress, applied stress, viscous term) are output for 1000 equidistant times during the run, which results in an effective time step $\Delta t\gg \delta t$. These data  are post-processed to extract dislocation positions and velocities. At each time step, the DPE is satisfied numerically with maximal absolute error of order $\smash{10^{-6}}$ \cite{JOSI18b}. Due to PBCs, $\siga(x,t)$ becomes uniform and steady once boundaries of the simulation box have been reached. Initial conditions consist of zero or one pre-existing dislocation at rest.

\begin{figure}[!ht]
	\centering
	\includegraphics[width=8.00cm]{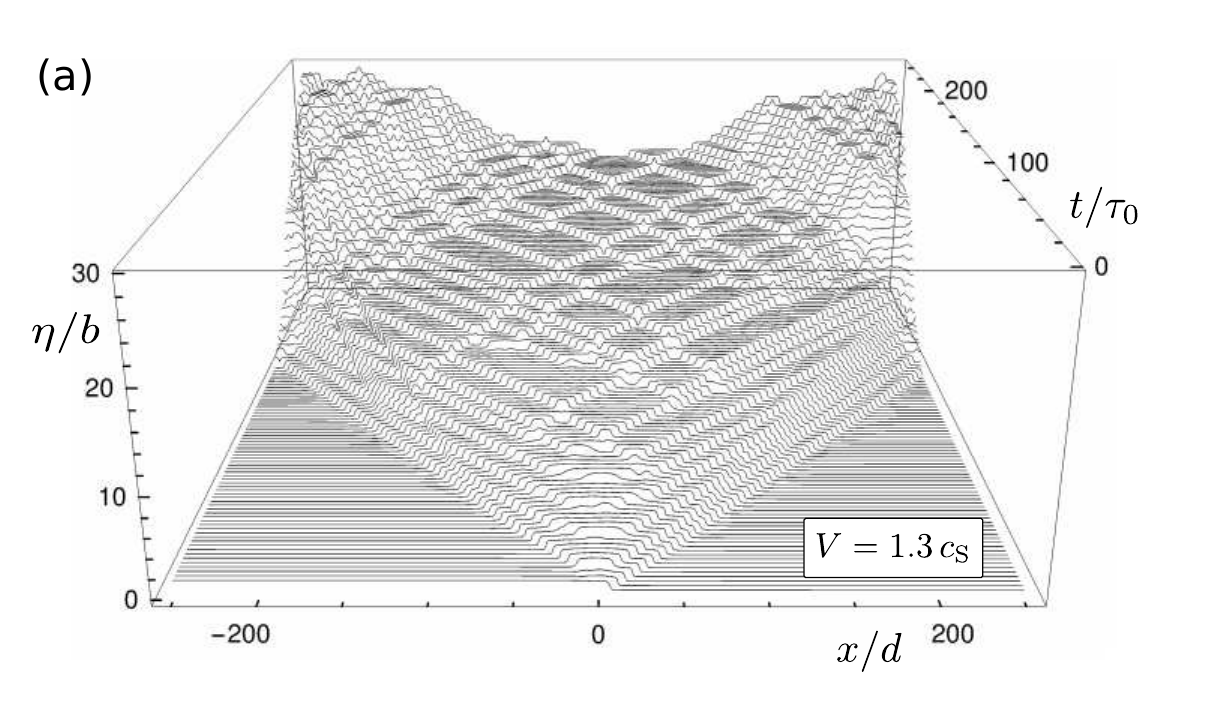}\\
	\includegraphics[width=8.00cm]{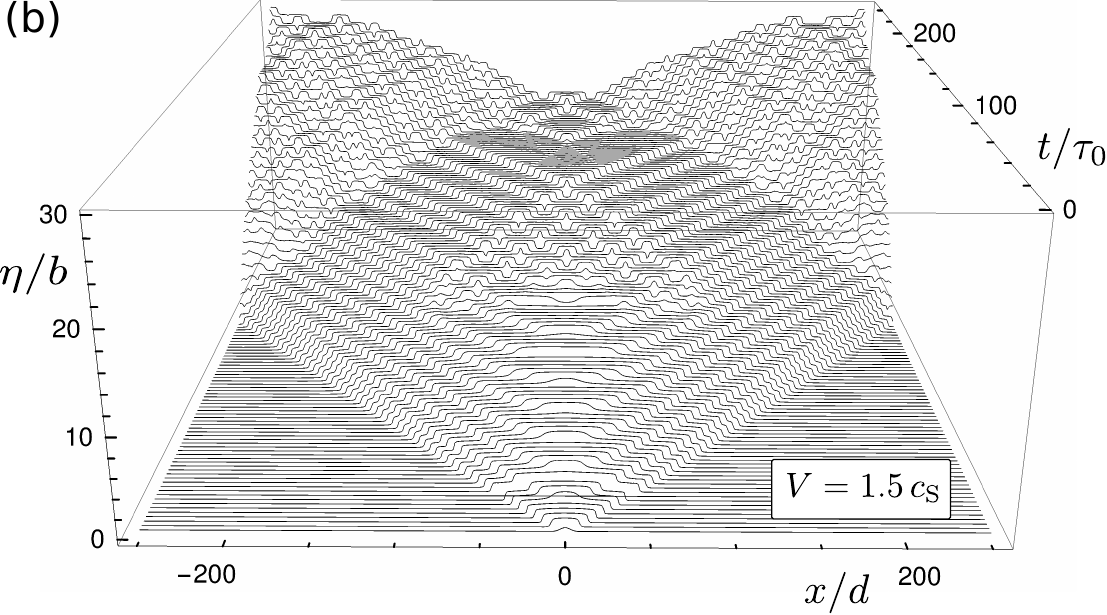}
	\caption{\label{fig:fig3} Edge dislocations. Evolution of $\eta(x,t)$ for $\siga=1.2\,\sigth$ (system size $L=160\,\pi$). (a) one initial dislocation at rest; (b) void initial state. `Shock' speeds $V$ as indicated.}
\end{figure}

Typical evolutions of $\eta(x,t)$ are represented in Fig.\ \ref{fig:fig3}. Hills at large times on both sides of the figures are due to collisions with reentrant dislocations. The asymmetry of the single-dislocation initial state in (a) is enhanced at large times through nucleation events.
\subsection{Results}
\subsubsection{Trajectories and overall system dynamics}
\label{sec:trajdyn}
Accounting for spatial discretization over $x$ of $\eta(x,t)$, individual dislocation trajectories \cite{WEIN75}  $\xi_i(t)$ ($i$ is a dislocation index) are identified by linear interpolation between discretization points as values $\xi_i$ of $\xi$ for which $|\eta(\xi,t)/b|$ is half-integer.
\begin{figure*}[t]
\centering
\begin{minipage}[c]{17.8cm}
    \includegraphics[width=17.4cm]{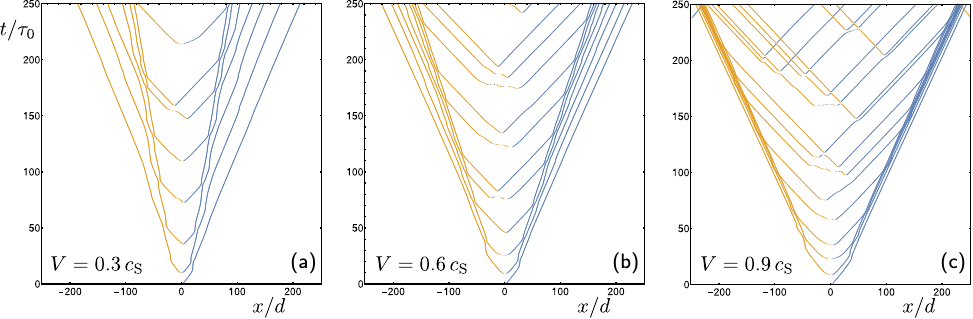}
	\medskip
	\includegraphics[width=17.4cm]{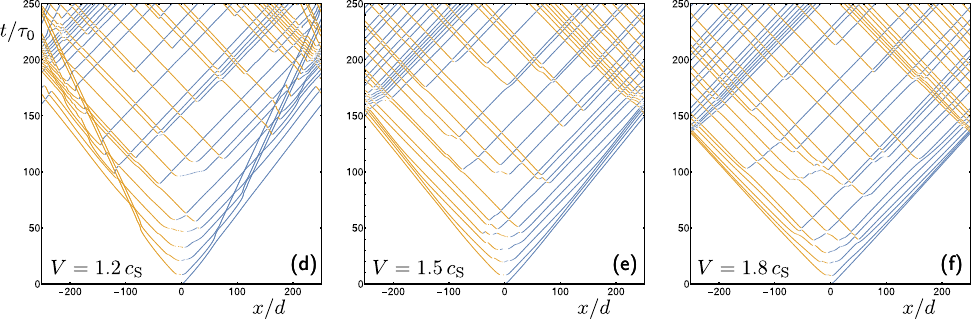}
	\medskip
	\includegraphics[width=17.4cm]{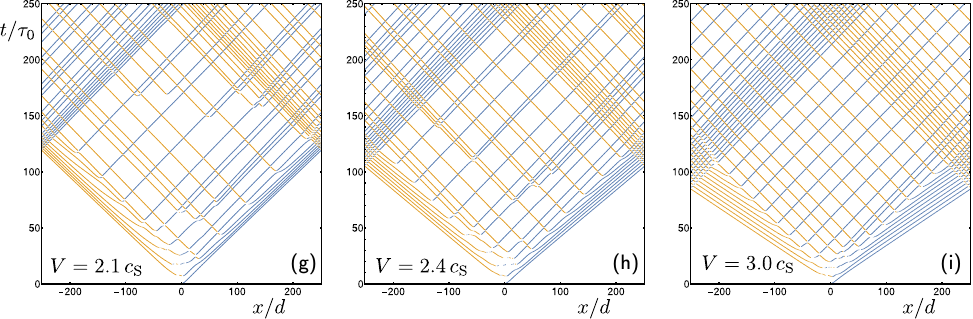}
	\caption{\label{fig:fig4} (Color online) Edge dislocation trajectories for $\max_x\siga(x,t)=1.2\,\sigth$, and various `shock' speeds $V$ as indicated, with one dislocation in the initial state. Negative (resp., positive) dislocations are in blue  (resp., orange).}
\end{minipage}
\end{figure*}
Figure \ref{fig:fig4} displays trajectories for $\siga=1.2\,\sigth$, and increasing values of $V$, for an initial state with one dislocation at rest (the smaller the slope of the trajectory, the larger the dislocation speed). 

The plots show that the dynamics starts off by sucessive nucleations of pairs near the center. This transient step is soon followed by nucleation by bursts in the bulk, with strong apparent randomness in positions and instants. It is emphasized that through the function $\eta(x,t)$, the Peierls model naturally embodies the variability of the local Burgers vector at nucleation \cite{AUBR11a}. Nucleations occur where the applied stress is maximal at the beginning of simulations, and afterwards preferentially near to pre-existing moving dislocations in reactions such as in Fig.\ \ref{fig:fig5} below, as was previously reported in a Frenkel-Kontorova model \cite{LOMD86}. Although the distance is much variable, this feature suggests a heterogeneous nucleation process, possibly involving nucleation avalanches \cite{BAGG23a} (but the system is too small to conclude on the latter point). Note, however, that the terminology \emph{heterogeneous nucleation}  \cite{SHEH06,*BEIG08,GUPT75,LI07} is usually reserved to nucleation near to pre-existing defects, other than previously-nucleated \emph{moving} dislocations.

The leading dislocations cross the box's boundaries via PBCs at time $t=t_{\text{BC}}$. Afterwards, the system is expected to enter a later steady regime (in a statistical sense) that has not been explored. In order to eliminate initial transients and the effect of PBCs, mean quantities will be extracted hereafter from data restricted to the time window $\smash{t\in T_{\text{av}}=[(2/3)t_{\text{BC}},t_{\text{BC}}]}$. A strong hypothesis of the forthcoming measurements of mean dislocation density and speeds is that within this window, the sytem achieves a state reasonably close to some steady-state. This might however be invalid due to finite-size and time effects. This potential limitation could be overcome by considering larger system sizes and simulation times.

\subsubsection{Reactions}
\begin{figure}[!ht]
	\centering
	\includegraphics[width=8cm]{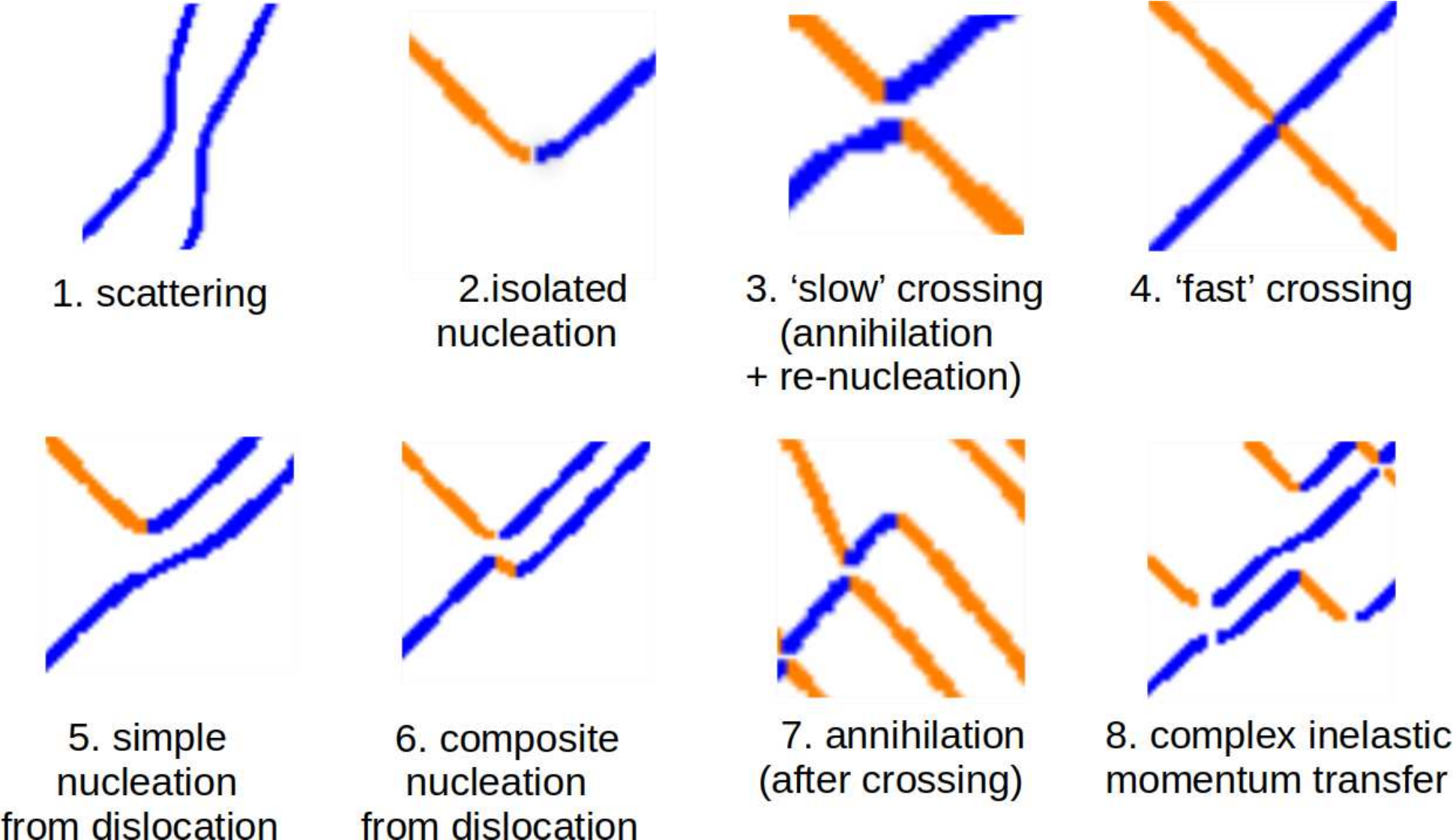}
	\caption{\label{fig:fig5} Dynamical reactions observed in calculations (blow-ups on trajectories such as in Fig.\ \ref{fig:fig4}, with same orientation and color coding).}
\end{figure}

The various reactions observed in Fig.\ \ref{fig:fig4} are categorized in Fig.\ \ref{fig:fig5}. The quasi-elastic reaction $\#$1 is observed at small $V$. Reactions $\#$2 to $\#$6 are the most frequent ones. Contrary to the frequent isolated nucleation $\#$2, annihilations are only observed in close connection with previous crossing events, $\#$7. However, the annihilation-scattering process at the center of $\#$8 (a time-reversed version of $\#$5) has not been observed individually. The apparent `pre-acceleration' of $\#5$ is explained by shape changes in the initial stages of pair nucleation. 

Dynamic collisions and crossings of dislocations of opposite signs and speeds occur via pair annihilation upon collision, and subsequent re-nucleation \cite{PILL09}. After annihilation, the available energy locally transferred to the elastic displacement field flows away with elastic wavespeed from the region of collision. In absence of applied stress the crossing can be completed by subsequent re-nucleation only if the dislocations possess enough initial momentum to re-create the pair before energy has escaped \cite{PILL09}. Here, with $\siga>\sigth$, re-nucleation just after collision is almost always guaranteed unless some perturbation reduces the local stress level for long enough a time to prevent it. This composite sequence (typically, $\#3$) may look quasi-instantaneous as in $\#4$, without appreciable differences between the `in' and `out' dislocations speeds (here, $v\simeq \cL$ for both). 

\subsubsection{Dislocation speeds}
\label{sec:speeds}
\begin{figure}[!h]
	\centering
	\includegraphics[width=8.4cm]{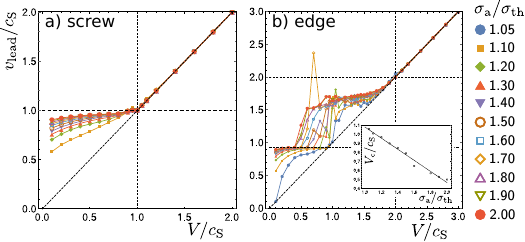}
	\caption{\label{fig:fig6} Speed $v_{\text{lead}}$ of leading dislocations versus `shock' speed $V$ (method M1), for various $\siga$. Inset in b): eye-estimated stress-dependent shock-speed threshold $V_c(\siga)$ for transition from subsonic to intersonic branch, versus stress $\siga$, with fit $V_c/\cS\simeq 1.70-0.62(\siga/\sigth)$.} 
\end{figure}
Once positions $\xi_i(t)$ have been determined as explained above, individual dislocation speeds $v_i(t)$ are estimated either as $\Delta \xi_i(t)/\Delta t$ (method M1), or by 
Eqs.\ \eqref{eq:parder} and \eqref{eq:locorowan} as $v_i(t)\simeq- \partial_t\eta(\xi_i(t),t)/\partial_x\eta(\xi_i(t),t)$ (method M2). Method M1 requires following the trajectories by continuity to perform the numerical time differentiation, which is problematic in presence of reactions such as above, except for isolated dislocations. Method M2 is free of the latter drawback, but can be inaccurate due to the unavoidable time variations of the dislocation cores embodied in its defining formula; see Eq.\ \eqref{eq:locorowangen}.

Of special importance is the speed $v_{\text{lead}}$ of the leading (i.e., outermost) dislocations evoked in Sec.\ \ref{sec:trajdyn} above. Speed $v_{\text{lead}}$, estimated with method M1 (the trajectory being unambiguous here), is displayed in Fig.\ \ref{fig:fig6} versus $V$ for screw and edge dislocations, for a range of values of $\siga/\sigth$. Speeds have been time-averaged over the interval $T_{\rm{av}}$. The outlier in plot b) is an artifact. The plot shows that $v_{\text{lead}}\geq V$ whatever $V$, with $v_{\text{lead}}=V$ for $V\geq \cS$ (resp. $\cL$) for screw (resp.\ edge) dislocations. Therefore, dislocations are always present ahead of or on the `shock' front. The observed low-stress dependence of the data for $v_{\text{lead}}<\cS$ is attributed to the dislocations not having reached their asymptotic subsonic velocity within the simulation window, due to slow relaxation \cite{PELL14}. Disregarding this, the data suggest that  $v_{\text{lead}}$ is quasi-independent of $\siga$, except for remarkable transitions of the edge dislocation from a subsonic branch of asymptotic states (with speed close to the Rayleigh velocity $\cR$) to an upper intersonic stable branch. This transition takes place at a stress-dependent shock-speed threshold $V_c(\siga)$ that decreases linearly as $\siga$ increases (see inset), which reveals a two-branch structure of $v_{\text{lead}}(V)$ akin to that of the stress/velocity curve of an edge dislocation (Fig.\ \ref{fig:fig1}). Both are most certainly related. In fact, the simulations reported in Fig.\ \ref{fig:fig4} suggest that a two-step loading process (see Sec.\ \ref{sec:dpe_overview}) takes place for $V<\cL$, as the leading edge dislocation is nucleated within the loaded zone with local stress $\smash{\sigma\simeq\sigth}$, but is afterwards expelled out of the latter into the unloaded zone ahead where it only experiences the weaker long-range stress of the train of dislocations behind.

Remarkably, although the present model admits no steady-state mobility law of a single edge dislocation for $\siga>\sigth$ and only allows for speeds restricted to the domain $v<\cL$,  dynamically-nucleated dislocations can nonetheless reach steady speeds $v>\cL$. Such speeds are not determined by $\siga$ but instead by the `shock' speed. So to say, leading dislocations are `pushed' forward by the dislocations continuously nucleated behind to accomodate shock advance.      

\begin{figure}[!ht]
	\centering
	\includegraphics[width=8.5cm]{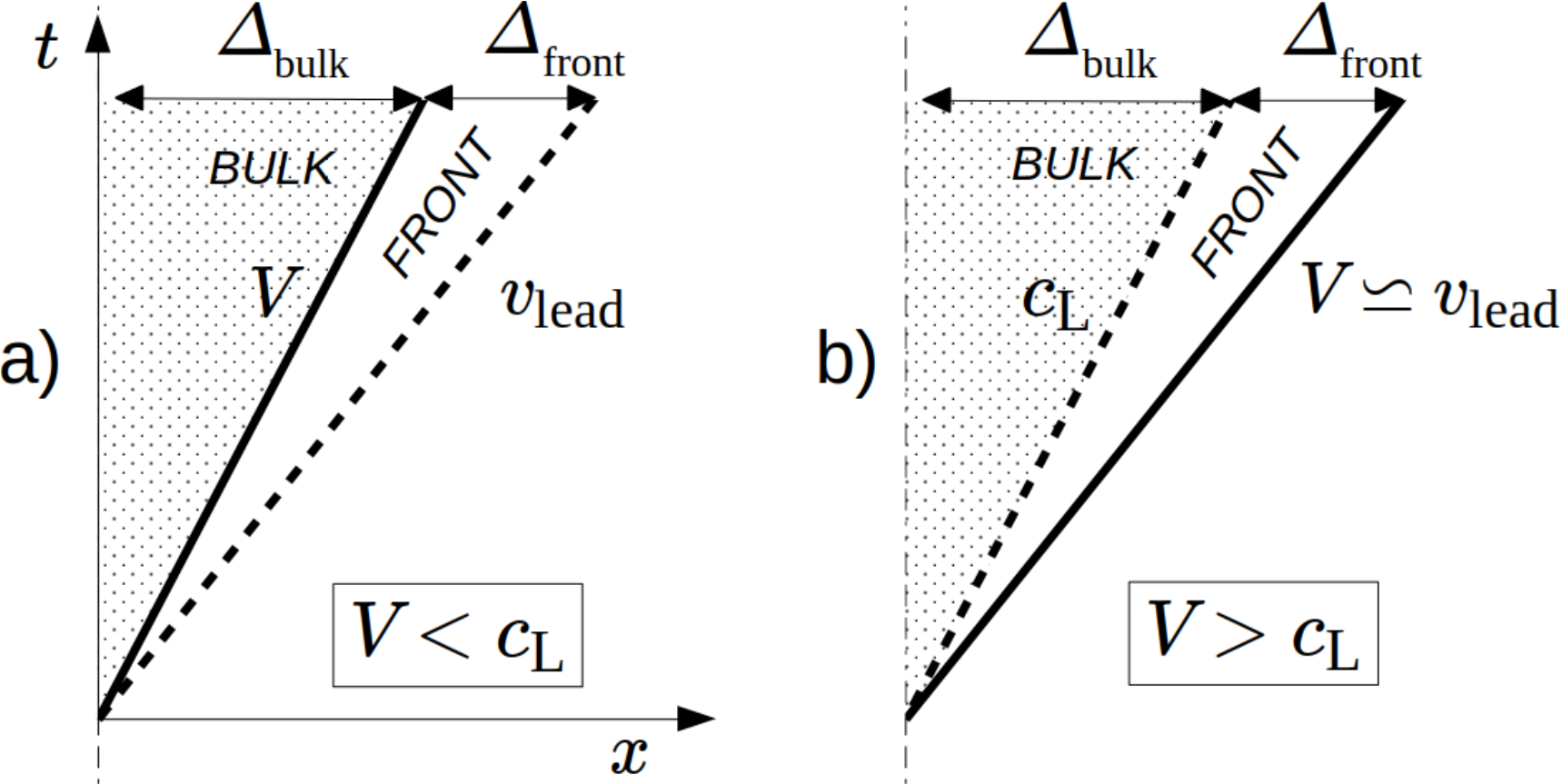}
	\caption{\label{fig:fig7} Definition of the \emph{Bulk} and \emph{Front} averaging zones in $(x,t)$ representation for an edge dislocation and for: (a) `shock' speed $V<\cL$; (b) shock-speed $V>\cL$. In both cases the stressed zone is $|x|< V\,t$. For a screw dislocation, $\cL$ should be replaced by $\cS$.}
\end{figure}
These differences of nature between the regimes $V\lessgtr\cL$ prompt us to introduce the notions of \emph{bulk} and \emph{front} zones, as depicted in Figs.\ \ref{fig:fig7} (a) and (b), of time-dependent widths $\Delta_{\text{bulk}}=\min(V,\cL)t$ and $\Delta_{\text{front}}=\Delta-\Delta_{\text{bulk}}$, where $\Delta(t)=v_{\text{lead}}\,t$ is the total extension of the process zone for $x>0$. The case $V>\cL$ figures an overdriven shock \cite{ZHAK11a}. 
\begin{figure}[!ht]
  \centering
  \includegraphics[width=8.4cm]{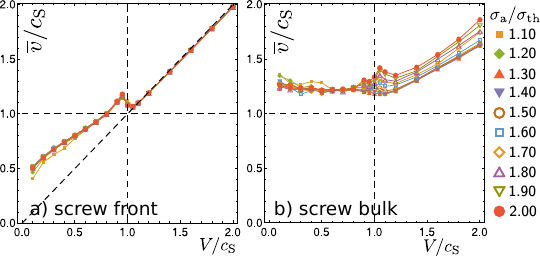}
  \caption{\label{fig:fig8}(Color online) Screw dislocations. Mean dislocation speed $\overline{v}$, Eq.\ \eqref{eq:meanvmeth2}, vs.\ shock speed $V$ for various $\siga$ (data sets in different colors) obtained from  averages over a) the front zone and b) the bulk zone.}
\end{figure}

\begin{figure}[!ht]
	\centering
    \includegraphics[width=8.4cm]{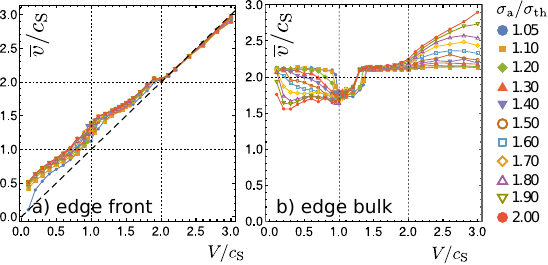}
	\caption{\label{fig:fig9}(Color online) Edge dislocations. Mean dislocation speed $\overline{v}$, Eq.\ \eqref{eq:meanvmeth2}, vs.\ shock speed $V$ for various $\siga$ (data sets in different colors) obtained from  averages over a) the front zone and b) the bulk zone.}
\end{figure}

The steady-state speed of the `bulk' dislocations after completion of the nucleation event (but not necessarily at larger times, due to reactions) can be understood from the following non-rigorous but intuitive argument. The absolute value of the local stress, $|\sigma(x,t)|$ is less than $\sigth$, but approaches that value in the majority of the bulk region,  as Fig.\ \ref{fig:fig11} below demonstrates. Then, the driving stress experienced by each dislocation of a newly nucleated pair is presumably close to $\sigth$, too. It can thus be substituted to $\siga$ in the stess-velocity law of Fig.\ \ref{fig:fig1}. Since for edges it is obviously larger than the  branch-selection threshold $\sigma_{\text{c}}$, the velocity branch selected is the upper one, with speed $\cL$, which is the maximal admissible one at $\siga=\sigth$. Likewise, the mobility law of a screw dislocation is single-branched, with speed $\cS$ at $\siga=\sigth$, which determines the nucleation speed of screws. The argument is made complete by considering that supersonic dislocations within each pair cannot interact via wave propagation, and can therefore be considered independent. Note that steady glide at the shear wavespeed of screw dislocations in dipole pairs was observed in atomistic simulations \cite{PENG19a}.

The dislocation speed $\overline{v}$ averaged separately over the front and bulk zones is represented versus $V$ in Fig.\ \ref{fig:fig8} (resp., \ref{fig:fig9}) for screw (resp., edge) dislocations. It is computed as
\begin{align}
\label{eq:meanvmeth2}
\overline{v}_z&=\av{\frac{1}{N_z(t)}\sum_{i\in z}|v_i(t)|}_{t\in T_{\rm{av}}}.
\end{align}
where the $v_i$ are estimated by method M2, the zone label $z$ is \emph{bulk} or \emph{front}, and $N_z(t)$ is the number of dislocations in the zone. In subplots (a) front dislocations move with mean speed increasing with shock speed $V$, dominated by a leading-order proportionality to the latter. Figure \ref{fig:fig9}(a) display weak overall dependence with respect to $\siga$, which is absent for screw dislocations in Fig.\ \ref{fig:fig8}(a), and might be a finite-size artifact. By contrast, the mean speed in the bulk in Fig.\ \ref{fig:fig9}(b) is markedly stress-dependent. This has two different causes: for $V<\cS$, the lowering of the mean speed at high stress is presumably due to the contribution of bulk dislocations nucleated close to $v=\cL$ that become subsonic after having `bounced' on the zone boundary. This effect, specfic to edge dislocations, becomes majoritary in the range $\cR<V<\cS$ whatever $\siga$, and is absent for screw dislocations in Fig.\ \ref{fig:fig8}(b). On the other hand, the high nucleation rate for $V>\cL$ increases the number of transient speeds $v\gg\cL$ sampled close to nucleation. It is present for both edge and screw dislocations.   

\begin{figure}[!htbp]
	\centering
	\includegraphics[width=8.4cm]{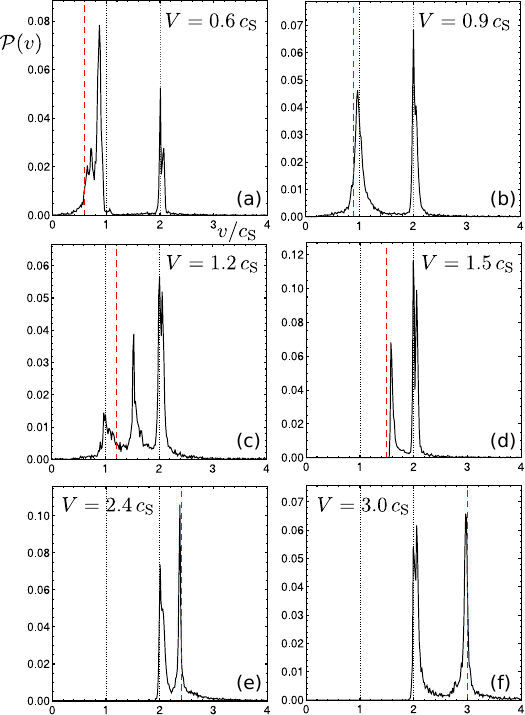}
	\caption{\label{fig:fig10}(Color online) Edge dislocations. Speed distributions $\mathcal{P}(v)$ for $\siga/\sigth$ $=1.2$, and shock speeds $V$ (dashed, red) as indicated}.
\end{figure}

Typical multimodal speed distributions for $\siga=1.2\sigth$, normalized to unity, collected over the whole simulation duration and over both regions, are displayed in Fig.\ \ref{fig:fig10} for edge dislocations. A peak close to $v=\cL=2.0\,\cS$ is always present, which corresponds to the `bulk' dislocations at their nucleation speed. By contrast, no large-amplitude peak is associated with the shock speed $V$ until $V>\cL$. Moreover, the system becomes populated with an appreciable amount of intersonic dislocations ($\cS<v<\cL$) as soon as $V$ exceeds the transition threshold $V_c(\siga=1.2\,\sigth)\simeq 0.96\,\cS$ (see inset of Fig.\ \ref{fig:fig6}). These intersonic dislocations form a well-defined peak if $\cS<V<\cL$. Quite strikingly, the system stands free of subsonic dislocations ($v<\cS$) if $V>1.3\,\cS$: then, \emph{plastic deformation essentially consist of intersonic/supersonic dislocations}. Finally, even for $V<\cL$ distributions display a small but appreciable amount of speeds $v\gg \cL$. Indeed, newly-nucleated dislocations, until fully formed, induce huge apparent measured transient speeds, as can be seen from trajectories in Fig.\ \ref{fig:fig4}. 
 
\subsubsection{Nucleation and local stress}
\label{sec:nucleation}
As recalled in Sec.\ \ref{sec:theeq} the athermal homogeneous nucleation threshold of a dipole identifies with the theoretical shear stress. Nucleation relieves the local excess stress in-between the two components of the nucleated dipole \cite{BRIN06,MEYE09}. As these components move away from one another (which would correspond to loop expansion in higher dimensions) this screening becomes less effective, until at some separation distance a new pair pops up to reinstate locally a condition $\smash{|\sigma|/\sigth\lesssim 1}$. 

\begin{figure}[!ht]
	\centering
	\includegraphics[width=8.4cm]{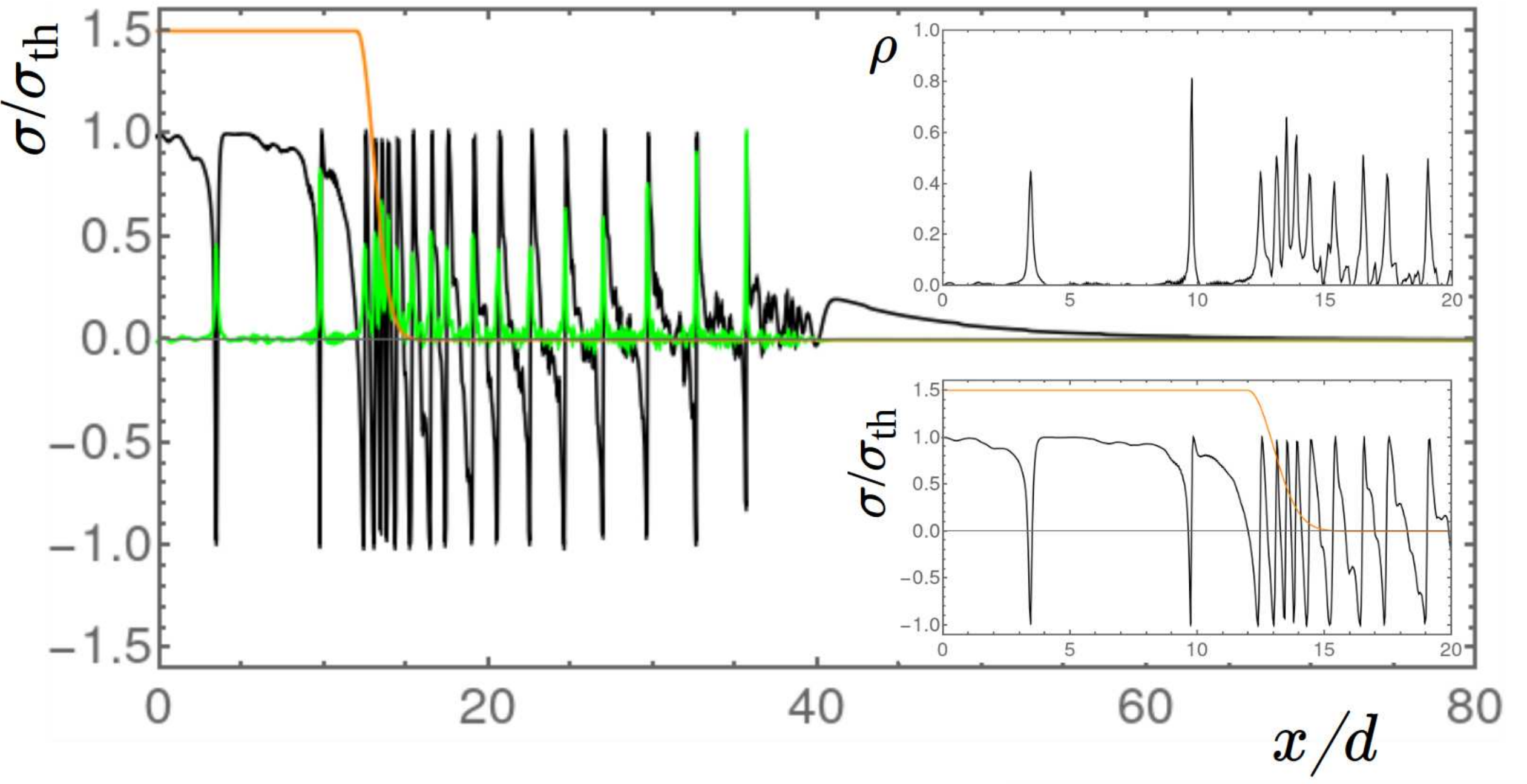}
	\caption{\label{fig:fig11} Edge dislocations. Main plot: at $t/\tau_0=39.93$, local stress $\sigma(x,t)$ for $\siga/\sigth=1.5$ and $V=0.30\,\cS$ (black), superimposed with simultaneous $\siga(x,t)$ (orange) and dislocation density $\rho(x,t)$  (green) -- divided by 50 for better representation. Top inset: blow-up of same dislocation density for $0\leq x/d\leq 20\times 2\pi$. Bottom inset: blow-up of same $\sigma(x,t)$ within this range.}
\end{figure}

Figure \ref{fig:fig11} displays, for subsonic $V$, a typical profile of $\sigma(x,t)$ at some intermediate time, together with the corresponding nucleated dislocation density (only half the system is shown). The local density $\rho$ being defined by Eq.\ \eqref{eq:parder}$_1$, the insets focused on the bulk/front boundary bewteen show that the stress and the dislocation cores are well resolved in our calculations. They also fluctuate, and stress heterogeneity \cite{GURR15e} is different ahead of and behind the shock front. The main plot evidences a forward stress tail. It arises in the form of a wave emitted by the first nucleated dislocation, and is afterwards reinforced in a cumulative way as more dislocations are nucleated behind. Its own front (unnoticeable on the figure) propagates with speed $\cL$. At the time represented, it is quasi-complete and its maximal amplitude does not change much thereafter. Similar pulses were observed in atomistic simulations \cite{ZHAK11a}. The strong shape variations of dislocation cores under the action of the surrounding wave system explains the difficulty of accurately estimating dislocation velocities with method M2 in the previous Section. 

Although the exact triggering events for nucleation could not be pinpointed, obvious candidates are fluctuations in the bath of elastic waves emitted and scattered by dislocations. The role of waves in `randomly' (i.e., pseudo-randomly) triggering nucleations is hinted at by considering cases $V>\cL$. Then the system entirely consists of supersonic dislocations with $v>\cL$ (see Fig.\ \ref{fig:fig9}). Elastic waves from these supersonic dislocations are radiated as Mach cones \emph{out of} the slip plane without any possibility of disturbing $\eta$. This makes interactions bewteen edge dislocations essentially local, so that a decrease in the number of `random' nucleation events is expected. This is precisely what Fig.\ \ref{fig:fig4}(i) illustrates at the highest shock speed $V=3\,\cS$ where a regular array of supersonic dislocations \cite{SMIT58,HORN62} is produced, to be investigated further below and in Appendix \ref{sec:sda}. Thus, for supersonic dislocations at very high shock speeds, nucleation turns out fully kinematic \cite{SCHI95,*SCHI02}.

\subsubsection{Dislocation spacings and densities}
Dislocation densities can be understood in terms of dislocation mean spacings. The overdriven case $V\geq \cL$ such as in Figs.\ \ref{fig:fig4}(g,h,i) is approximately addressed remarking that the front zone consists of a regularly-spaced array of supersonic dislocations of same sign moving with speed $v\simeq V$. Based on the Weertman equation, a model of dislocation spacing $\Delta x(v,\tau)$ in the array is derived in Appendix \ref{sec:sda}, leading to the dislocation spacing in the front zone
\begin{align}
\label{eq:deltaxf}
\Delta x_{\text{f}}(V,\tau)&=(2\pi)d\,B_\alpha(V)/\sqrt{\tau^2-1}\qquad (V\geq\cL),
\end{align}
where $B_\alpha(v)$ is the drag coefficient at supersonic speed $v$ and $\tau=\siga/\sigth$ is the reduced stress.    
\begin{figure}[!ht]
	\centering
	\includegraphics[width=8.4cm]{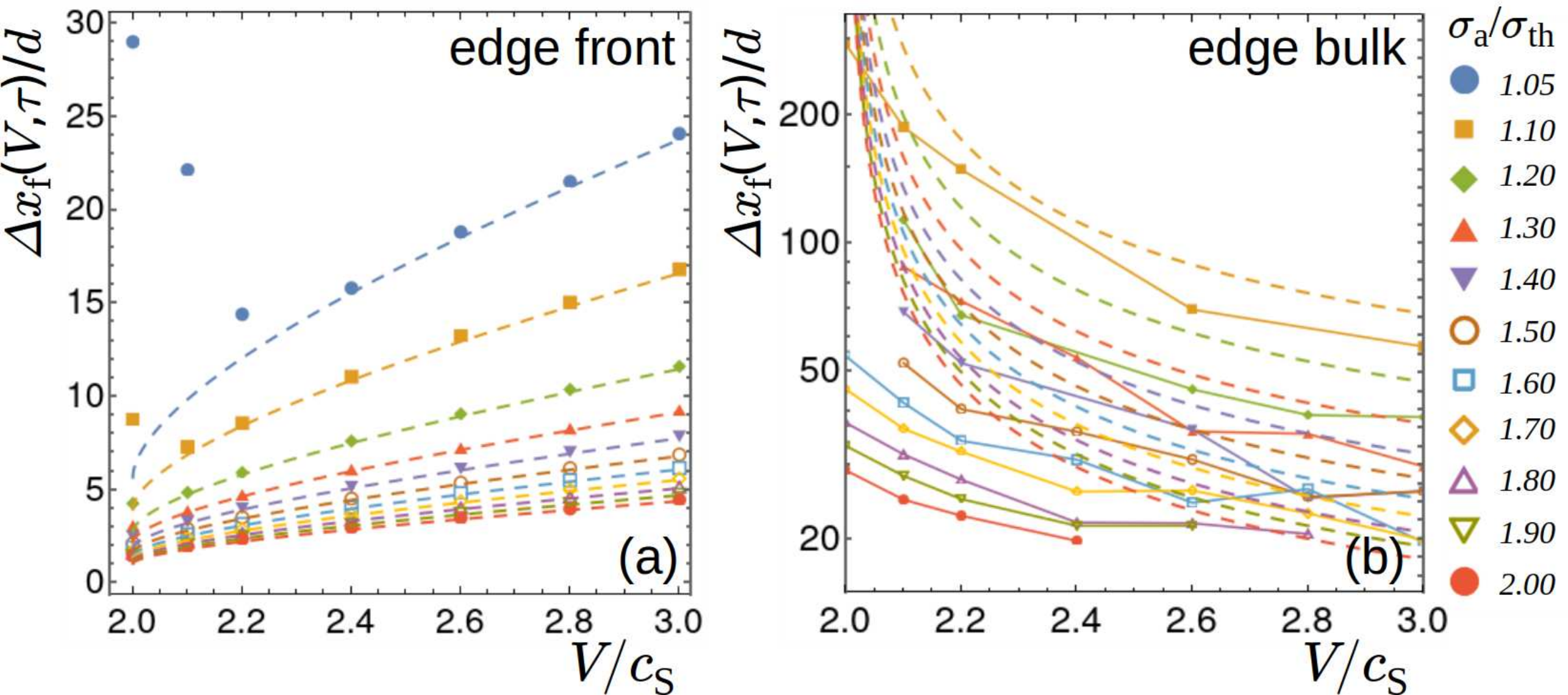}
	\caption{\label{fig:fig12} For $V\geq\cL$, dislocation spacings vs.\ $V$, for reduced stresses $\tau=1.05, 1.1, 1.2,\ldots, 2.0$ from top (blue) to bottom (red), in the front zone (a) and the bulk zone (b, in log.\ scale). Markers:  measurements. Dashed lines: Eqs.\ \eqref{eq:deltaxf} in (a) and \eqref{eq:deltaxb} in (b).}.
\end{figure}

Figure \ref{fig:fig4} suggests that for $V\gg\cL$ dislocation pairs are almost exclusively nucleated at the bulk/front boundary, with one dislocation going into the bulk zone and the other one into the front zone. Under this assumption, the nucleation rate (number of dislocations nucleated per unit time) for each zone is the relative speed $\Delta v$ at which dislocations move away from the boundary, divided by the mean dislocation spacing. This speed being $\Delta v=2\cL$ in the bulk zone, and $\Delta v=V-\cL$ in the front zone, equating nucleation rates $\Delta v/\Delta x$ gives the following estimate for the mean spacing for dislocations of same sign in the bulk zone as
\begin{align}
\label{eq:deltaxb}
\Delta x_{\text{b}}(V,\tau)&\simeq\frac{2\cL}{V-\cL}\Delta x_{\text{f}}(V,\tau)\qquad (V\gg\cL).
\end{align}

In Fig.\ \ref{fig:fig12} measurements of spacings between dislocations \emph{of same sign}, averaged over over a few outermost dislocations in the front zone, and over a few dislocations in the bulk, are compared to Eqs.\ \eqref{eq:deltaxf} and \eqref{eq:deltaxb}. Figure (a) shows excellent overall agreement with \eqref{eq:deltaxf}, in spite of deviations at small $V$ for the lowest $\siga$s, which indicate that the steady-state has not been reached there. In (b) by constrast, Eq.\ \eqref{eq:deltaxb} gives a correct order of magnitude only when $V\gg\cL$, due to the crude nature of our argument (data relative to parameter values for which a relevant spacing could not be determined due to excessive reaction-induced perturbations have been omitted in the plots). Indeed, an ordered `steady state' in the bulk for $V>\cL$ consists of \emph{two} arrays of dislocation of opposite signs steadily moving in opposite directions, which periodically annihilate one another (e.g., Fig.\  \ref{fig:fig4} (i)). Such a pattern is intrinsically dynamic and escapes the Weertman equation used to derive \eqref{eq:deltaxf}.

Using Eq.\ \eqref{eq:parder}, the mean (unsigned) dislocation density $\overline{\rho}$ is estimated via measurements in bulk and front zones as 
\begin{align}
\overline{\rho}_{\text{z}}&=\av{\int_{\text{z}}\dd x\,|\rho(x,t)|/\Delta_{\text{z}}(t)}_{T_{\text{av}}},
\end{align}
where the zone label z is \emph{front} or \emph{bulk}. The integral is evaluated as a Riemann sum. A further average is then made over time within the useful time window.

For $V>\cL$, Eqs.\ \eqref{eq:deltx} and \eqref{eq:rhobar} in Appendix \ref{sec:sda} suggest a steady-state mean density in the front zone
\begin{align}
\overline{\rho}_{\text{f}}(V,\tau)&=(b \Delta x_{\text{f}})^{-1}\propto\sqrt{\tau^2-1}.
\end{align}
This estimate does not compare quantitatively well with the data and is not considered further. This may be due to various causes among which dislocation core shapes varying much stongly than in the theoretical model (see Fig.\ \ref{fig:fig9}). 

However, our data are consistent with a generalized scaling factor in stress of the form $\smash{S_\beta(\tau)=(\tau^2-1)^\beta}$ with $0<\beta<1$ an empirical complexity-related zone-dependent rescaling exponent.
\begin{figure}[!ht]
	\centering
	\includegraphics[width=8.5cm]{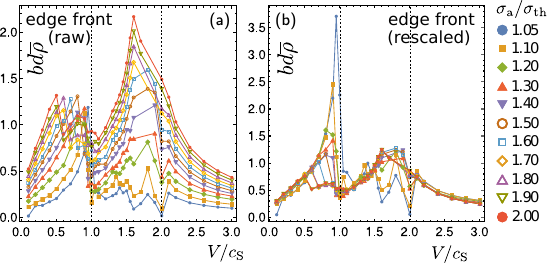}
	\caption{\label{fig:fig13} (Color online) Edge dislocations. (a) Mean density $\overline{\rho}$ in the front zone vs.\ $V$ for various stresses and (b) data collapse with rescaling exponent $\beta=1/2$.}
\end{figure}
Figure \ref{fig:fig13}(a) displays measured densities in the front zone for edge dislocations. Apart from deviations attributed to finite-size effects, the rescaling in (b) is consistent with an exponent $\beta=1/2$ over the whole range of $V$. The transitions for $V<\cS$ in Fig.\ (b) should be brought together with those in Fig.\ \ref{fig:fig6}.
\begin{figure}[!ht]
	\centering
	\includegraphics[width=8.5cm]{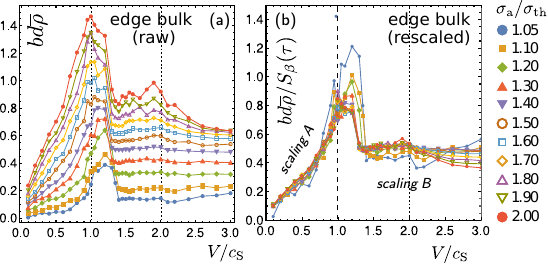}
	\caption{\label{fig:fig14} (Color online) Edge dislocations. (a) Mean density $\overline{\rho}$ in the bulk zone vs.\ $V$ for various stresses and (b) data collapse with different rescaling exponents $\beta=3/4$ in region A, and $\beta=1/2$ in region B.}
\end{figure}
Mean density data in the bulk zone are reported in Figure \ref{fig:fig14}(a), while in (b) exponents were needed for $S(\tau)$ in regions A ($V<\cS$) and B ($V>\cS$) as indicated in the caption. See Supplemental Material \cite{SUPP} for data similar to Figs.\ \ref{fig:fig13} and \ref{fig:fig14} for screw dislocations. 

\section{Concluding discussion}
\label{sec:concl}
Within a 1D model, we investigated high-stress multiple nucleations of dislocation pairs on a single slip plane under (elasto)dynamic conditions by means of an original method. Although the framework is free of side complexities such as Peierls stress or forest hardening \cite{MADE02b}, the outcome turns out surprisingly rich.

With regard to dislocation-based constitutive models, the study suggests that the notion of a local dislocation-density nucleation rate (such as usually considered in models) may not straightforwadly apply to an advancing shock. Indeed, in absence of forest interactions, dislocations are continuously and kinematically nucleated so as to maintain a quasi-steady density of geometrically-necessary dislocations \cite{ARSE99} that screens out the (macroscopic) applied stress. This is much alike the findings of Zhakhovsky \textit{et al.}\ in atomistic simulations \cite{ZHAK11a}, and of Gurrutxaga-Lerma \textit{et al.} in subsonic discrete dislocation elastodynamic simulations \cite{GURR15a}. Of relevance as well is the observation that whatever this applied stress, the inner stress that drives dislocations and determines their speed via the mobility law \emph{should never exceed the nucleation threshold}. A distinction between the applied stress and this local stress should therefore presumably be included in models.

	Applying the model to a highly idealized `shock' situation, we found the system to  spontaneously separate into two 'bulk' and 'front' zones. Whereas Meyers \cite{MEYE78}, opposing Smith's views \cite{SMIT58}, considered dislocation generation rather than motion responsible for the accomodation of lattice deviatoric strains, the present work suggests that both generation and motion are equally important in shocks: here dislocations are generated only either in the bulk zone or at the interface between the bulk and front zone, at supersonic speed, and can be expelled into the nucleation-free front zone. In particular, supersonic `shock' speeds cause supersonic dislocations to transit from the bulk to the front zone in this model.

An accurate theory is presently lacking to account for the densities reported in Figs.\ \ref{fig:fig13} and \ref{fig:fig14} over the whole domain of parameters.
However, dislocation densities were found to scale as $((\siga/\sigth)^2-1)^\beta$, where $\siga$ is the applied stress, $\sigth$ is the theoretical shear stress (nucleation stress), and $0<\beta<1$. Exponent $\beta$ does not depend on the shock speed $V$ in the front zone, where it is of order $\beta\sim 0.5$. In the bulk zone, it differs depending on whether $V<\cS$ ($\beta\sim 0.75$ for edges, and $\sim 0.66$ for screws) or $V>\cS$ ($\beta\sim 0.5$ for both characters). These values should however be reexamined on larger systems for better statistics. Moreover, they may depend on the $\gamma$-potential, since the base form $[(\siga/\sigth)^2-1]^{1/2}$ (Appendix \ref{sec:sda}) is very much related to the sine Frenkel form. A scaling of the type $[(\siga/\sigth)^{\beta_1}-1]^{\beta_2}$ with two variable exponents might however hold more generally. Because the PND equation includes no non-trivial equation of state (EoS), in spite of some similarities the present two-zone picture differs from the one discovered by Zhakhovsky \textit{et al.}\ in the overdriven regime \cite{ZHAK11a}. One major difference is that our front zone is continuously expanding, but this can be due to 1D steric effects and may not occur in higher dimensions.

Our model's approximations may affect the results in at least three aspects. First, our framework assumes linear elasticity, except for the pull-back force. However, real shocks involve nonlinear elasticity due to the non-trivial EoS. This would result in pressure-dependent elastic constants and higher wave speeds behind the shock front compared to ahead of it. However, in compressed metals, Poisson's ratio remains approximately constant in the absence of phase transformation. Consequently, the wavespeed ratio $\gamma=\cL/\cS$ in the kernels presented in Appendix \ref{sec:kernels} would remain unchanged by compression \cite{KANE04}. Hence, uniformly compressed stress steps should yield similar outcomes within the loaded region (encompassing both the "bulk" and "front" zones) with same dislocation reactions. On the other hand, if the wavespeeds in the loaded "bulk" zone are higher than those in the unloaded "front" zone for the case of $V<\cL$, the self-organization of the front might change.

Second, the underlying physics of the DPE may not be complete. Introducing local inertia in the pull-back force, as discussed in Section \ref{sec:theeq}, would likely alter the dynamics and introduce additional delays in nucleation and reactions.

Third, in our 1D simulations, all dislocations turn out mobile. Overcoming this limitation would require expanding the analysis to higher dimensions to account for junctions and other reactions \cite{GURR15a,GOME06}. These three issues require detailed investigations that extend beyond the scope of this paper.
	
Nevertheless, in spite of these possible limitations, the present study embodies full inertia of radiative origin \cite{PELL12,PELL14}, drag, wave propagation and retarded interactions on the slip plane, nucleation and annihilations, and supersonic motion, within one self-contained field equation without the need for extra inputs. Therefore, it might serve as a one-dimensional, reference point for linear-elastic, discrete elastodynamic simulations, as well as a tool to evaluate future nucleation-related equations for dislocation density evolution.	  

\acknowledgments
Y.P.P.\ thanks his colleague C.\ Denoual for discussions and a remark about reactions.

\begin{appendix}
\section{Kernels and their transforms}
\label{sec:kernels}
The various kernels needed are as follows (the climb-edge \cite{WEER67a} case is reported for completeness only, as it is not used in our simulations). Kernels $K_a(x,t)$ \cite{PELL10b,PELL12} vanish for $t<0$ and are given below for $t>0$ together with their associated constants $\kappa_a$. Their spatial Fourier transforms (FT) $K_a(k,t)$, of wavemode $k$, are expressed in terms of functions $C_\alpha(\tau)$, which also vanish for $t<0$, as 
\begin{align}
\label{eq:conversionKC}
K_a(k,t)&=-\ii\frac{\pi}{2}\cS k\,C_a(\cS k t).
\end{align}
Bearing in mind the usual correspondence between dislocation characters and fracture modes, the $C_a(\tau)$ are reproduced hereafter from Refs.\ \cite{COCH97,*GEUB95} in terms of the wavespeed ratio $\gamma=\cL/\cS>1$, and of the integral
\begin{align}
\label{eq:wdef}
W(\tau)&=1-\int_0^\tau \frac{\dd q}{q}J_1(q),
\end{align}
which admits a closed-form expression \footnote{\protect  Indeed, $\int_0^x\dd y\,J_1(y)/y=\frac{\pi x}{2}\left[J_1(x)\mathbf{H}_0(x)-J_0(x)\mathbf{H}_1(x)\right]+x J_0(x)-J_1(x)$, where $\mathbf{H}_\nu$ is Struve's function. This follows from formula 6.511.6 in I.S.\ Gradshteyn and I.M.\ Ryzhik, \emph{Table of Integrals, Series and Products}, 7th.\ ed. (Academic Press, Amsterdam, 2007) and the identity $J_0(x)=J_1(x)/x+J_1'(x)$.}. Finally, the numerical method of solution (see Appendix \ref{sec:methsol}) needs the Laplace transforms (LT) 
\begin{align}
\label{eq:lapdef}
\mathcal{C}_a(s)&=\int_0^{+\infty}\dd\tau\,C_a(\tau)\ee^{-s\tau},
\end{align}
which we give here in a slightly reorganized form with respect to
\cite{COCH97,*GEUB95} to prepare for Eq.\ \eqref{eq:CL} below.
\begin{widetext}
	Thus for $t>0$, 
	\begin{subequations}
		\label{eq:kernels}
		\begin{align}
		\label{eq:Ks}
		\text{screw}:\qquad  
		&K_{\text{s}}(x,t)=\frac{1}{2\cS}\frac{x}{t^2}(\cS^2 t^2-x^2)_+^{-1/2}
		\qquad\text{with}\qquad \kappa_{\text{s}}=1,\\
		\label{eq:cstau}
		&C_{\text{s}}(\tau)=\frac{J_1(\tau)}{\tau},\\
		\label{eq:css}
		&\mathcal{C}_{\text{s}}(s)=\sqrt{1+s^2}-s,\\
		\label{eq:Kg}
		\text{glide edge}:\qquad
		&K_{\text{g}}(x,t)=\frac{2\cS^2}{x^3}\left[\frac{1}{\cL}(2\cL^2 t^2-x^2)(\cL^2 t^2-x^2)_+^{-1/2}-\frac{1}{\cS}(2\cS^2 t^2-x^2)(\cS^2 t^2-x^2)_+^{-1/2}\right]\nonumber\\
		&\hspace{1cm}{}+\frac{x}{2\cS t^2}(\cS^2 t^2-x^2)_+^{-1/2}+\frac{\cS}{2}x\Pf(\cS^2 t^2-x^2)_+^{-3/2} \qquad\text{with}\qquad \kappa_{\text{g}}=1,\\
		\label{eq:cgtau}
		& C_{\text{g}}(\tau)=\frac{J_1(\tau)}{\tau}
		+4\tau\left[W(\gamma \tau)-W(\tau)\right]-\frac{4}{\gamma}J_0(\gamma \tau)+3J_0(\tau),\\
		\label{eq:csg}	&\mathcal{C}_{\text{g}}(s)=-\frac{4}{s^2}\left[\sqrt{1+(s/\gamma)^2}
		-\frac{(1+s^2/2)^2}{\sqrt{1+s^2}}\right]-s,\\
		\label{eq:Kc}
		\text{climb edge}:\qquad
		&K_{\text{c}}(x,t)=
		-\frac{2\cS^2}{x^3}\left[\frac{1}{\cL}(2\cL^2 t^2-x^2)(\cL^2 t^2-x^2)_+^{-1/2}-\frac{1}{\cS}(2\cS^2 t^2-x^2)(\cS^2 t^2-x^2)_+^{-1/2}\right]\nonumber\\
		&\hspace{1cm}{}+\frac{\cL}{2\cS^2}\frac{x}{t^2}(\cL^2 t^2-x^2)_+^{-1/2}+\frac{\cS^2}{2\cL}\left(\frac{\cL^2}{\cS^2}-2\right)^2 x\Pf (\cL^2 t^2-x^2)_+^{-3/2}
		\qquad\text{with}\qquad \kappa_{\text{c}}=\gamma,\\
		\label{eq:cctau}
		&C_{\text{c}}(\tau)=\gamma^3\frac{J_1(\gamma \tau)}{\gamma \tau}
		+4\tau \left[W(\tau)-W(\gamma\tau)\right]+(4\gamma-\gamma^3)J_0(\gamma \tau)-4 J_0(\tau),\\
		\label{eq:csc}		&\mathcal{C}_{\text{c}}(s)=-\frac{4}{s^2}\left[\sqrt{1+s^2}-\frac{(1+s^2/2)^2}{\sqrt{1+(s/\gamma)^2}}\right]- \gamma s,
		\end{align}
	\end{subequations}
\end{widetext}
where $\Pf$ denotes Hadamard's finite part and $(x)_+^\alpha=x^\alpha$ if $x>0$ and $0$ otherwise \cite{GELF64}. Equations \eqref{eq:Ks} and \eqref{eq:Kg} are, respectively, equations (7a) and (7b) of \cite{PELL12}, written differently. Equation \eqref{eq:Kc} derives from Eqs.\ (43a) and (43b) of \cite{PELL10b} and the transformation explained in \cite{PELL12}. Letting $s=\ii v/\cS$, both terms within brackets in \eqref{eq:csg} and \eqref{eq:csc} are proportional to the Rayleigh function. Their common root $v^*=\cR$ is the Rayleigh wavespeed of free-surface waves \cite[p.\ 18]{RAVI04}, equal to $\cR\simeq 0.9325\,\cS$ if $\cL=2\,\cS$ as in the present work.

Quite remarkably, expressions \eqref{eq:css}, \eqref{eq:csg}, and \eqref{eq:csc} show Laplace transforms $\mathcal{C}_a(s)$ to be directly connected to the so-called prelogarithmic steady-state Lagrangian functions \cite{BELT68,*HIRT98,PELL12,*PELL20} $L_a(v)$ for each character $a$, with $v$ the dislocation velocity, via the relationship ($w_0=\mu b^2/(4\pi)$ is the characteristic line energy density)
\begin{align}
\label{eq:CL}
\mathcal{C}_a(s)&=-\frac{1}{w_0}L(\ii\cS s)-\kappa_a\,s.
\end{align}
This connection went previously unnoticed to the best of our knowledge. Moreover, the constant $\kappa_a$ can be obtained in all cases from $L(v)$ as
\begin{align}
\kappa_a=-\frac{1}{w_0}\lim_{s\to\infty}\frac{1}{s}L(\ii\cS s),
\end{align}
which is the same limit as Eq.\ (66) of \cite{PELL12}, formulated otherwise.
\section{Numerical method}
\label{sec:methsol}
\label{sec:fastobli}

The main difficulty of \eqref{eq:dpe2} lies in the integrodifferential convolution operator, which is non-local in time and space. Nevertheless, this operator is diagonalized by the Fourier transform in space, followed by a Laplace transform in time. Our method crucially relies on this property. Although used in a different way (notably, with respect to the time convolution) those transforms are central as well in the approach of Rice and co-workers in the elastodynamic crack problem, which moreover involves the same elastodynamic kernels \cite{COCH97,*GEUB95}; see Appendix \ref{sec:kernels}.

In the rest of the Section, the substitution $\cS t\to t$ is implied to simplify notations, so that time has dimension of space. Using Eq.\ \eqref{eq:conversionKC} the DPE reads in Fourier form
\begin{align}
\label{DPN_k}
\kappa^\alpha_a\partial_t \delta\eta(k,t)+k^2\int_{-\infty}^t\hspace{-1em}  \dd t' C_a\bigl(|k|(t-&t')\bigr) \delta\eta(k,t')\\
&=-(2/\mu)F(k,t).\nonumber
\end{align}
The kernels involved are stiff in space and time, because of their proportionality to $k^2$, and because the kernel $C_a(|k|(t-t'))$ oscillates faster and faster as $k\to\infty$. The simulation interval of the $x$-axis is discretized with step $\Delta x$ and the continuous FT is approximated by the Discrete FT using FFT routines. For simplicity of exposition, the discrete character of the Fourier modes is left implicit hereafter.

In statics, the fast Fourier transform (FFT) technique has been employed repeatedly to solve the PN equation \cite{XIAN08,JOSI18a}.
Since here all functions are smooth, Gibbs oscillations are not a problem and no filtering or modification of the wavemodes is required. However, use of the DFT makes $\delta\eta$ periodic in space. No attempt has been made to mitigate this drawback, which could partially be overcome using zero-padding \cite{PRES97,JOSI18b}; see also \cite{COCH97}.

Then, time integration requires first re-expressing \eqref{DPN_k} as 
\begin{align}
\label{DPN_kres}
\delta\eta(k,t)&=-\frac{2}{\mu}\int_0^t \dd t'\,R_a\bigl(|k|(t-t')\bigr)F(k,t'),
\end{align}
where the resolvent $R_a$ appears as the Fourier transform of the \emph{fundamental solution}  $g_a$ -- more precisely $g_a(k,t)=-(2/\mu)\mathcal{R}_{\text{a}}(|k|t)$. 
The latter is the solution of the instance of \eqref{eq:dpe2} associated with a point source $F(x,t)$ $\to$ $\smash{F^{\text{point}}}(x,t)=\delta(t)\delta(x)$.
The advantage of \eqref{DPN_kres} over \eqref{DPN_k} is that the stability properties of the convolution operator, which are hidden in the oscillations of $C_a$, appear more explicitly as a long-range decay of $R_a$ \cite{JOSI18b} (this has beneficial numerical consequences.)

Then, for time integration, we make use of Lubich and Sch\"adle's so-called \emph{fast and oblivious} method \cite{LUBI02,*SCHA06}, where the adjective \emph{oblivious} actually indicates that the information storage of the past times is `parsimonious'.
This method proved the most efficient one in terms of accuracy, speed, and memory requirements, among the wide variety of strategies thoroughly tested and compared for the present problem by the second author (M.J.) \cite{JOSI18b}.
The main ideas of the method, which rests on a numerical inversion of the Laplace transform, are as follows.

Eq.\ \eqref{DPN_kres} is solved by a LT as
\begin{align}
\label{PDN_kp}
\delta\eta(k,s)&=
-\frac{2}{\mu}\frac{1}{|k|}\mathcal{R}_{\text{a}}(s/|k|)\mathcal{F}(k,s),
\end{align}
where $\mathcal{F}(k,s)$ is the LT of $F(k,t)$, and where 
\begin{align}
\label{eq:reslap}
\mathcal{R}_{\text{a}}(s)&=\frac{1}{\kappa^\alpha_a s+\mathcal{C}_a(s)}
\end{align}
is the LT of the resolvent
\begin{align}
\label{eq:resolvent}
R_a(\tau)&=\int_{\Gamma_\text{B}}\frac{\dd s}{2\ii \pi}\mathcal{R}_{\text{a}}(s)\,\ee^{s\tau},
\end{align}
where $\Gamma_\text{B}$ is a Bromwich inversion contour.
To evaluate \eqref{DPN_kres}, the continuous integral in the Laplace inversion of $\mathcal{R}_\text{a}$ must be done numerically in discretized form. Knowledge of the poles and cuts in the complex $s$-plane of $\mathcal{R}_\text{a}$ is necessary. (These functions can be expressed so that their cuts lie horizontally parallel to the real axis.) 
Following Lubich, Sch\"adle and co-workers, the Bromwich contour $\Gamma_\text{B}$ is deformed into a set of non-overlapping Talbot contour(s) $\Gamma_l$ with $l$ an integer index, suitably parametrized to enclose from the right the poles and branch cuts.
Being confined within horizontal strips of the complex plane, Talbot contours are well-suited to this inversion problem.

Integral \eqref{eq:resolvent} is then approximated by a Riemann sum of $d$ exponential terms, as
\begin{align}
\label{eq:approxR}
R_a(|k|t)&\simeq 
\sum_{j=1}^d \frac{w_j}{2\ii\pi}  \mathcal{R}_{\text{a}}(s_j) \ee^{s_j|k|t},
\end{align}
where $\{s_j\}_{j=1,\ldots,d}$ is a $|k|t$-dependent discrete sample of points on the set of contours, and the $w_j$ are parametrization-de\-pen\-dent weights for trapezoidal integration.
We employed the discretization parametrization of Ref.\ \onlinecite{LUBI02}, which enjoys spectral approximation error.
Typically, the number $d$ is not too large ---of order $100$-$500$ for accuracy better than $10^{-8}$ on $R_a$; see Ref.\ \onlinecite[Fig.\ 8.7]{JOSI18b}. However, exponentials in \eqref{eq:approxR} explode for $t$ large when $\Re(s_j)>0$. To ensure stability quantities $\Re(s_j)|k|t$ are kept under control by making contours tighter around singularities as time grows, whenever $|k|t$ falls into specific intervals of values.

Substituting \eqref{eq:approxR} into \eqref{DPN_kres} gives
\begin{subequations}
	\begin{align}
	\label{eq:detau}
	\delta\eta(k,t)&\simeq \sum_{j=0}^d w_j u_j(k,t),
	\end{align}
	with for $j=1,\ldots,d$,
	\begin{align}
	\label{eq:ujkt}
	u_j(k,t)&\equiv-\frac{2}{\mu}\frac{\mathcal{R}_{\text{a}}(s_j)}{2\ii\pi} 
	\int_0^t \dd\tau\,\ee^{s_j |k|(t-\tau)}F(k,\tau).
	\end{align}
\end{subequations}
Further differentiating \eqref{eq:ujkt} with respect to time then yields evolution equations for the $u_j(k,t)$ as
\begin{align}
\label{eq:ode}
\frac{\dd u_j}{\dd t}(k,t) 
&=s_j |k| u_j(k,t)-\frac{2}{\mu}\frac{\mathcal{R}_{\text{a}}(s_j)}{2\ii\pi} F(k,t).
\end{align}
Thus, the integral equation \eqref{DPN_k} has been replaced by a simpler system  of $d$ first-order evolution equations. This approach avoids keeping in memory the $F(k,\tau)$ for $0<\tau<t$, in favor of advancing the $u_j(k,t)$ in time for each $s_j$. A hierarchical handling of the past information on the contour-related intervals of $|k|t$ confers the algorithm its excellent scaling properties in time. We refer to the original publications \cite{LUBI02,*SCHA06} and to Ref.\ \cite{JOSI18a} for details on the (delicate) procedure for selecting the suitable contour at each time step.
In practice, Eqs.\ \eqref{eq:ode} are discretized in time by means of the so-called \emph{Radau II A} Runge-Kutta method of order $\bigO{\Delta t^5}$, with a fixed time step $\Delta t$ common to all modes $k$. Its stability \cite{HAIR96} makes this method particularly advantageous to address the stiffness of Eq.\ \eqref{eq:ode}. If $T$ is the number of time steps, and $M$ is the number of points in the spatial grid the algorithm has a complexity of $\bigO{d M\log M\times T\log T}$ and requires a memory of $\bigO{dM\log T}$, to be compared with $\bigO{M^2 T^2}$ and $\bigO{MT}$, respectively, for a naive evaluation of the time convolution in \eqref{DPN_k}. 

\emph{To summarize:} starting from initial conditions $\eta_0(x)$, and $u_j(k,0)=0$ for each discrete wavemode $k$, and given: the expressions of $\mathcal{R}_{\text{a}}(s)$; sets of discretization points $s_j$ on suitable Talbot contours; and a model for $\siga(x,t)$, the algorithm consists in looping over the following tasks for each time step $t_k$: (i) compute the elastic slip $\etae(x,t_k)$ associated with $\siga$; (ii) evaluate $F(x,t_k)$ in direct space as \eqref{eq:Fdef}; (iii) going to the Fourier domain, compute $F(k,t_k)$ by FFT; (iv) evolve for each discrete wavemode $k$ the $u_j(k,t)$ to the next time step $t_{k+1}$ by means of \eqref{eq:ode}; (v) reconstruct $\delta\eta(k,t_{k+1})$ from the $u_j(k,t_{k=1})$ using \eqref{eq:detau}, and deduce $\delta\eta(x,t_{k+1})$ by inverse FFT. At $t_k$ the plastic slip is $\eta(x,t_k)=\eta_0(x)+\delta\eta(x,t_k)$.

\section{Supersonic dislocation array}
\label{sec:sda}
The following applies to the front region, for $V>\cL$. Once divided by $\sigth$, and upon setting $\tau=\siga/\sigth$, Weertman's equation \cite{ROSA01,PELL14} reduces, in the supersonic regime $|v|>\cL$ for edges and $|v|>\cS$ for screws, to 
\begin{align}
\label{eq:wss}
2\pi \frac{d}{b}B_\alpha(v)\frac{\partial\eta}{\partial x}(x)+\tau&=\sin\left(\frac{2\pi}{b}(\eta(x)+\eta_e)\right)
\end{align}
in the co-moving frame of speed $v$. Here, $B_\alpha(v)$ is 
the radiative drag coefficient augmented by its empirical phonon-drag contribution of dimensionless viscosity parameter $\alpha>0$, and $\eta_e$ is the background elastic slip \cite{PELL14}. Because of the supersonic regime, the integral term of the full Weertman equation \cite{ROSA01} is absent from \eqref{eq:wss}.
Letting $\beta_{\text{L}}$  $=\sqrt{v^2/\cL^2-1}$, $\beta_{\text{S}}=\sqrt{v^2/\cS^2-1}$, and $s_v=\sign{v}$, the supersonic radiative drag coefficient reads
\begin{subequations}
\begin{align}
B^{\text{edge}}(v)&=2 s_v\left(\frac{\cS}{v}\right)^2
\left(\beta_{\text{L}}+\frac{\left[1-v^2/(2\cS^2)\right]^2}{\beta_{\text{S}}}\right),\\
B^{\text{screw}}(v)&=\frac{s_v}{2}\beta_{\text{S}},
\end{align}
\end{subequations}
and $B_\alpha(v)=B(v)+\alpha v/\cS$. 

In the strong-stress case $|\tau|>1$, $\eta_e=(b/4)\sign{\tau}$ \cite{PELL14} and  Eq.\ \eqref{eq:wss} admits a multiple-dislocation solution akin to a Smith array of dislocations of speed $v$. Eshelby only briefly evoked its existence, describing it as ``a linear function of $x-v t$ with a superimposed periodic ripple'' \cite{ESHE56a}. Its master analytical form was given by Movchan et al.\ \cite{MOVC98b}, albeit in a different context that did not require the function $B_\alpha(v)$ (see the latter reference for a plot). It can be built from the following base solution of \eqref{eq:wss} indexed by $k\in\mathbb{Z}$:
\begin{align}
&\eta^\text{base}_k(x)=-\frac{b}{4}\sign{\tau}+\frac{b}{\pi} 
\arctan\biggl\{\frac{1}{\tau}\nonumber\\
&{}-\frac{\sqrt{\tau^2-1}}{\tau}
\tan\left[\pi\frac{x}{\Delta x(v,\tau)}-Q(\tau)\right]\biggr\}-k\,b,
\end{align}
where the integration constant
\begin{align}
Q(\tau)&=\arctan\sqrt{(|\tau|-1)/(|\tau|+1)}
\end{align}
has been adjusted so that $\eta^\text{base}_{k=0}(0)=0$. The function $\smash{\eta^\text{base}_k}$ is bounded, periodic, and discontinuous at points
\begin{align}
x_k^{\text{disc}}&=\Delta x(v,\tau)\left[\left(k-\frac{1}{2}\right)+\frac{1}{\pi}Q(\tau)\right],
\end{align}
but with continuous derivative. Its spatial period, which corresponds to the distance between dislocations in the array, is 
\begin{align}
\label{eq:deltx}
\Delta x(v,\tau)&=\frac{2\pi|B_\alpha(v)|d}{\sqrt{\tau^2-1}}.
\end{align}
The array solution itself stems from assembling pieces of $\eta^\text{base}_k$ in a continuous way, as
\begin{align}
\label{eq:staircase}
\eta^{\text{array}}(x)&=\eta^\text{base}_{[\![(x-x^{\text{disc}}_0)/\Delta x(v,\tau)]\!]}(x),
\end{align}
where the value of $k$ in $\eta^\text{base}_k$ results from taking an integer part (denoted by $[\![\cdot]\!]$), which involves $\smash{x^{\text{disc}}_{k=0}}$. The mean dislocation density in the array is, as expected,
\begin{align}
\label{eq:rhobar}
\overline{\rho}&=\int_0^{\Delta x}\frac{\dd x}{\Delta x}\,\left|\frac{\partial \eta^{\text{array}}}{\partial x}\right|=\frac{1}{\Delta x(v,\tau)}.
\end{align}
\end{appendix}

\onecolumngrid
\widetext
\clearpage
\begin{center}
	\textbf{\large Supplemental Material: Shock-driven nucleation and self-organization of dislocations in the dynamical Peierls model
	}
	\medskip
	
	Y.-P. Pellegrini${}^{1,2}$ and M. Josien${}^3$
	\medskip
	
	\emph{
		${}^1$ CEA, DAM, DIF, F-91297 Arpajon, France\\
		${}^2$ Universit{\'e} Paris-Saclay, LMCE, 91680 Bruy{\`e}res-le-Ch{\^a}tel, France\\
		${}^3$ CEA Cadarache, F-13108 St-Paul-lez-Durance, France
	}
\end{center}
\setcounter{equation}{0}
\setcounter{figure}{0}
\setcounter{table}{0}
\setcounter{section}{0}
\setcounter{page}{1}
\makeatletter
\renewcommand{\theequation}{S\arabic{equation}}
\renewcommand{\thefigure}{S\arabic{figure}}
\renewcommand{\thetable}{S\arabic{table}}
\renewcommand{\thesection}{S~\Roman{section}}
\section{Mean-field collective-variable approximation vs.\ full numerical solution}
\label{sec:meanfield}
This section presents, for one edge dislocation, comparisons aimed at cross-checking the numerical solution of the Dynamical Peierls Equation (DPE) against its mean-field collective-variable-approximation (CVA) of Ref.\ [36] in the main text (hereafter referred to as (I)\,). 

\begin{figure}[!ht]
	\centering
	\includegraphics[width=4.3cm]{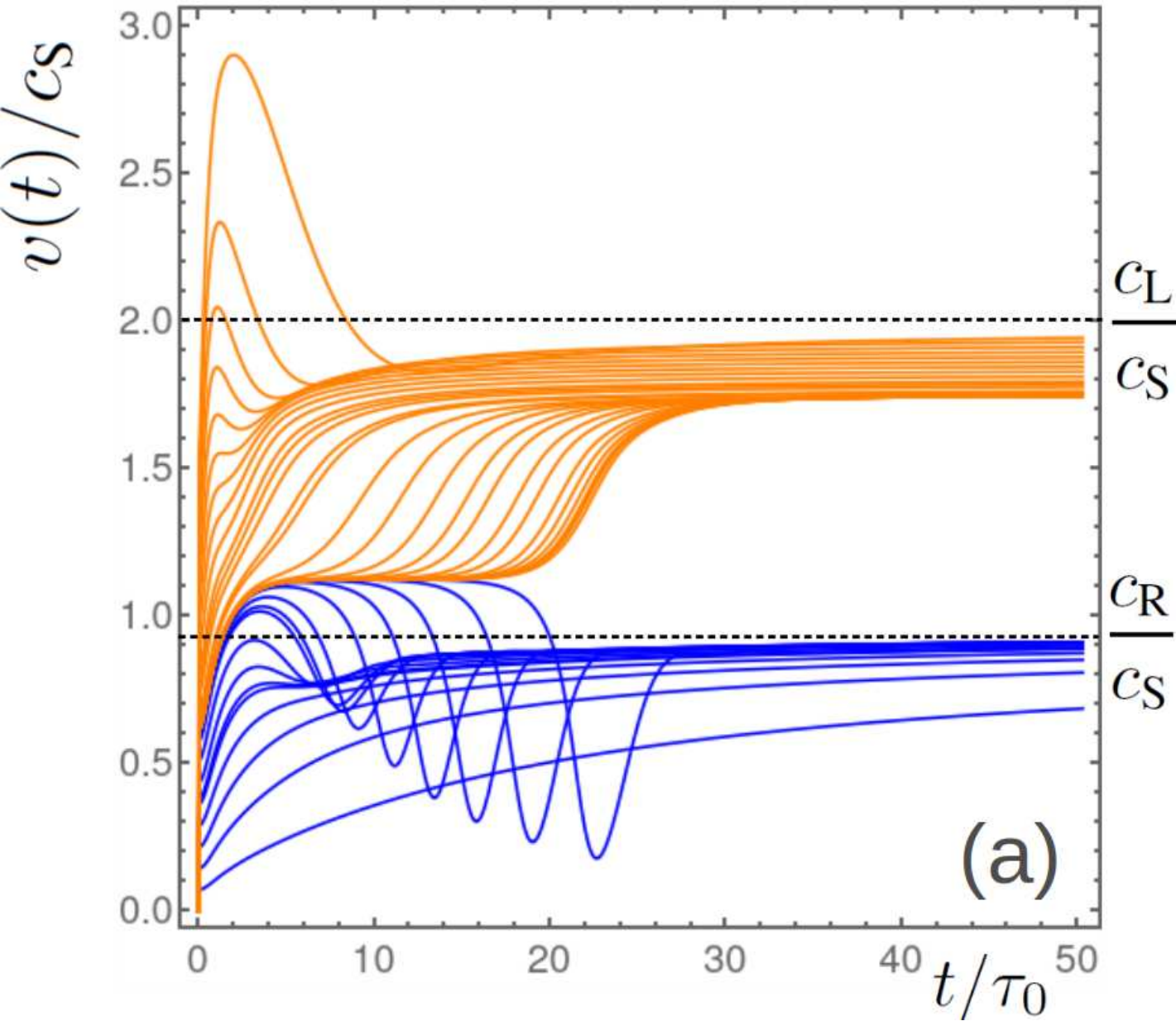}
	\includegraphics[width=4.1cm]{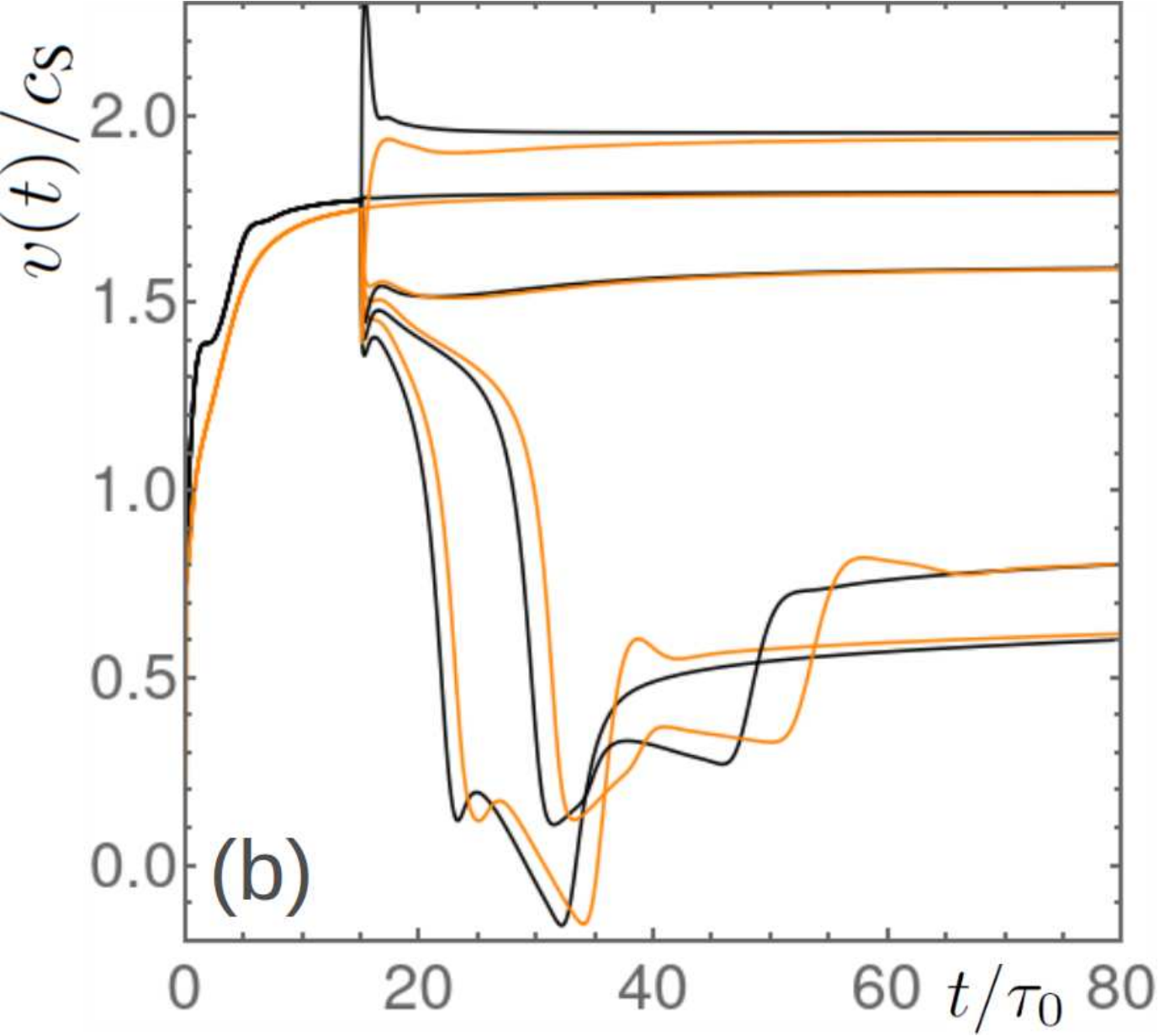}
	\caption{\label{S_fig:fig1} Edge dislocation speed $v(t)$ vs.\ time $t$, for: (a) single-step loading, with: (blue) loadings $\siga<\sigma_{\rm c}$; (orange) loadings $\siga>\sigma_\text{c}$, where $\sigma_\text{c}\simeq 0.401\sigth $; (b) double-step loading, black: CVA, orange: present full solution. Compare with Fig.\ 3 and Fig.\ 11c of (I).}
\end{figure}
Figure \ref{S_fig:fig1}(a) displays the dislocation speed vs.\ time for various single-step loadings $0.048\leq \siga/\sigth\leq 0.952$ from the full numerical solution. Comparing with Fig.\ 3 of (I), the overall behavior is qualitatively similar, with same asymptotic states (a.s.) (see Fig.\ 1 in the main text). 

Time-dependent inertia of radiation-reaction origin is less [s1,s2] than inertia approximated with velocity-dependent masses [s3].
Figure \ref{S_fig:fig1}(a) further shows that the full numerical solution experiences more inertia and damping than the CVA, as illustrated by Figs.\ \ref{S_fig:fig1}(a,b): unlike the CVA in (I) the full solution is fully overdamped, with extra delay in the response to loading, and in the transitions. All else being equal, the CVA underestimates radiative damping since it relies on a fixed master core shape, whereas the large number of degrees of freedom of the full solution offers a larger number of dissipation routes.

In Fig.\ \ref{S_fig:fig1}(b), analogous to Fig.\ 11(c) of (I) for double-step loading, both solutions are superimposed, with reasonable match. The finding of (I) that the asymptotic state strongly depends on the applied loading is thus confirmed by the DPE.

\begin{figure}[!ht]
	\centering
	\begin{minipage}[c]{10cm}
		\includegraphics[width=4cm]{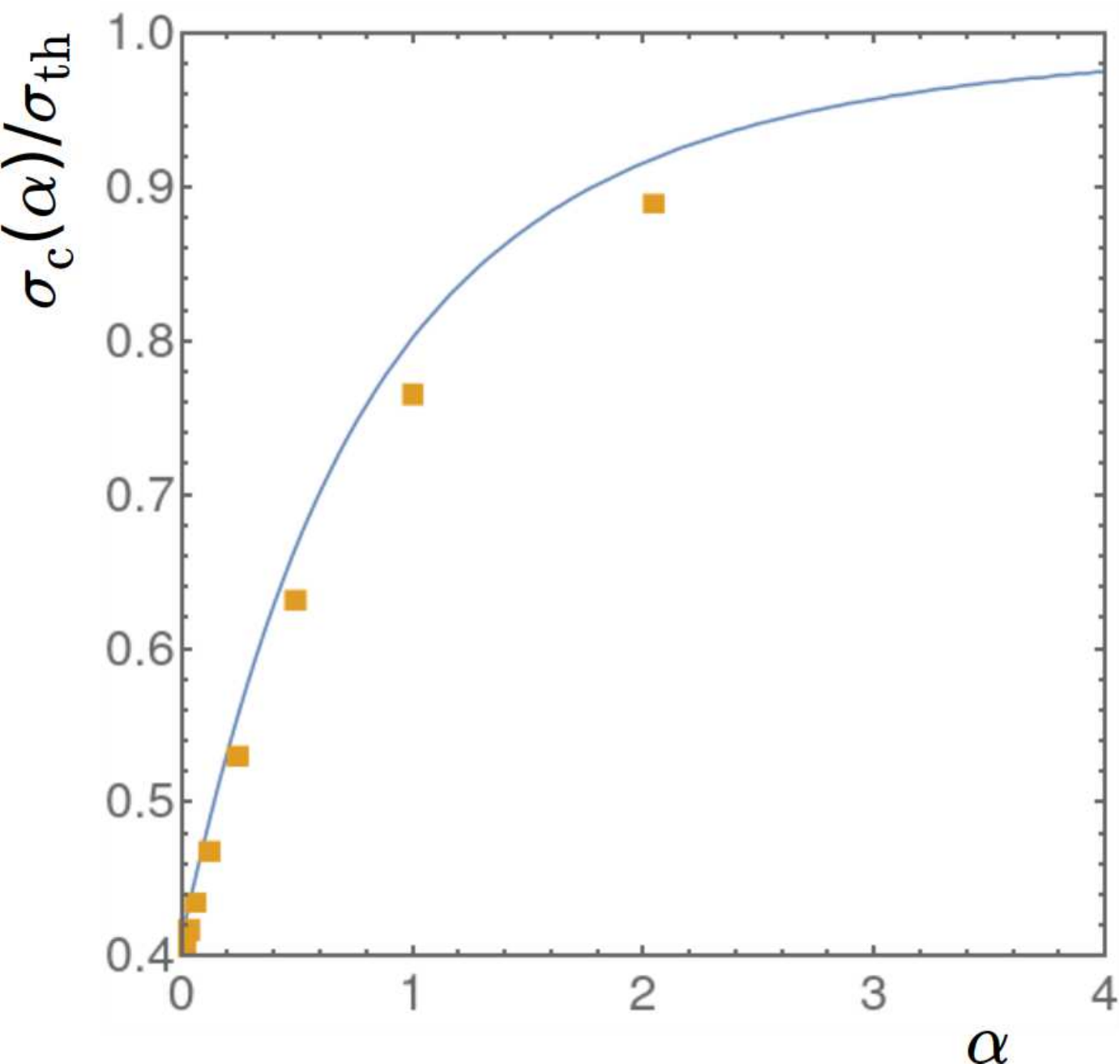}
		\caption{\label{S_fig:fig2} Critical stress $\sigma_\text{c}(\alpha)$ vs.\ phonon drag coefficient $\alpha$. Solid: CVA theory; orange markers: full numerical solution. Compare with Fig.\ 5 of (I).}
	\end{minipage}
\end{figure}

As in (I), and for a phonon-drag coeffcient $\alpha=0.01$, a critical loading $\sigma_\text{c}\simeq 0.401\,\sigth$ separates subsonic from intersonic asymptotic states. Although slightly lower, this value is close to the CVA value $\sigma_\text{c}\simeq 0.415\,\sigth$ reported in (I). The same holds for larger values of $\alpha$; see Fig.\ \ref{S_fig:fig2}.
Therefore, in spite of slight differences, overall good agreement between the dynamics in the full numerical solution of the DPE and in its CVA is found. Specifically, the CVA predictions regarding the subsonic-intersonic transition of edge dislocations in the DPE are confirmed.

\section{Density data for screw dislocations}
Density data for screw dislocations similar to Figs.\ 13 and 14 in the main text are as follows. 
\begin{figure}[!ht]
	\centering
	\includegraphics[width=8.5cm]{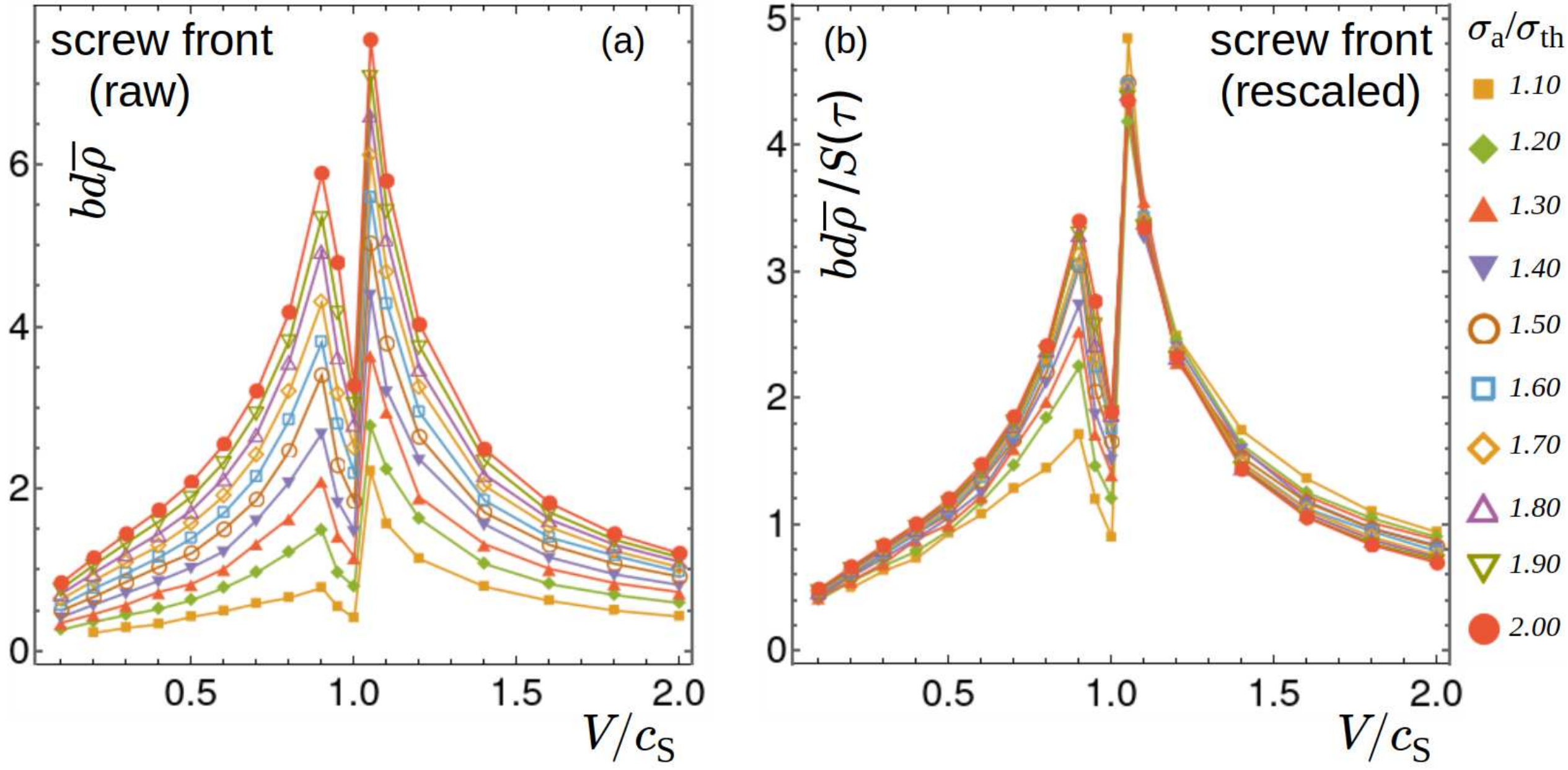}
	\caption{\label{S_fig:fig3} Edge dislocations. (a) Mean density $\overline{\rho}$ in the front zone vs.\ $V$ for various stresses and (b) data collapse with rescaling exponent $\beta=1/2$.}
\end{figure}
Figure \ref{S_fig:fig3}(a) displays mean densities in the front zone. The rescaling in (b) is consistent with an exponent $\beta=1/2$ over the whole range of $V$.
\begin{figure}[!ht]
	\centering
	\includegraphics[width=8.5cm]{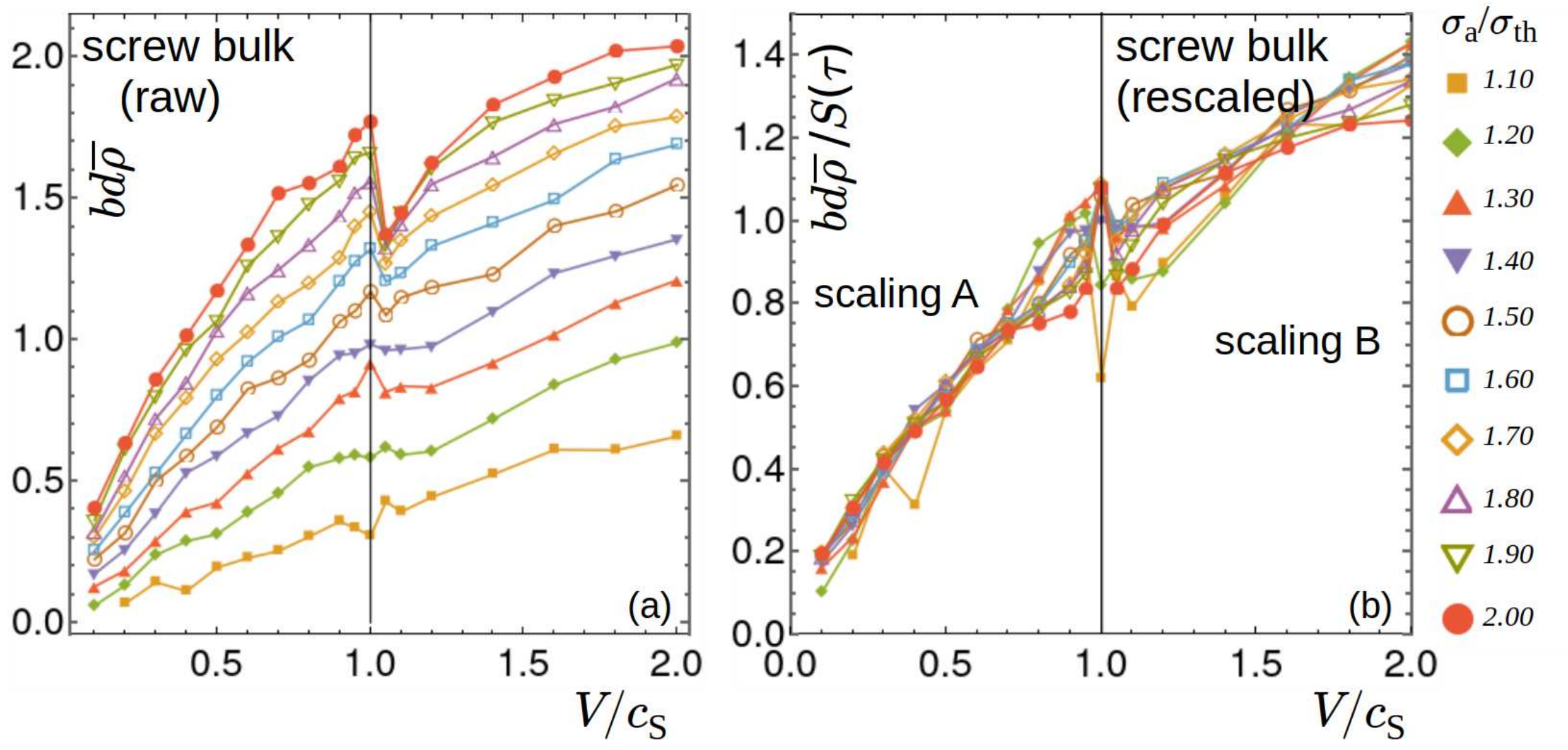}
	\caption{\label{S_fig:fig4} Edge dislocations. (a) Mean density $\overline{\rho}$ in the bulk zone vs.\ $V$ for various stresses and (b) data collapse with different rescaling exponents $\beta=0.66$ in region A, and $\beta=0.45$ in region B.}
\end{figure}
Mean density data in the bulk zone are reported in Figure \ref{S_fig:fig4}(a), while in (b) different exponents were needed for $S(\tau)$ in regions A ($V<\cS$) and B ($V>\cS$) as indicated in the caption. As in the edge case, `bulk' data thus needs two different exponents for $V<\cS$ and $V>\cS$.

\section{Singularities of Laplace transform of the resolvent}
This section addresses the issue of the singularities -- poles of cuts -- in the complex plane of the Laplace transform (B4) of the resolvent, namely,
\begin{align}
\mathcal{R}_{\text{a}}(s)&=\frac{1}{\kappa^\alpha_a s+\mathcal{C}_a(s)}.
\end{align}
The necessary assumption for the numerical Laplace inversion technique described in Appendix B of the main text is that these singularities all lie in the half-plane $\Re(s)\leq 0$. 

\begin{figure}[!ht]
	\centering
	\begin{minipage}[c]{16cm}
		\includegraphics[width=5cm]{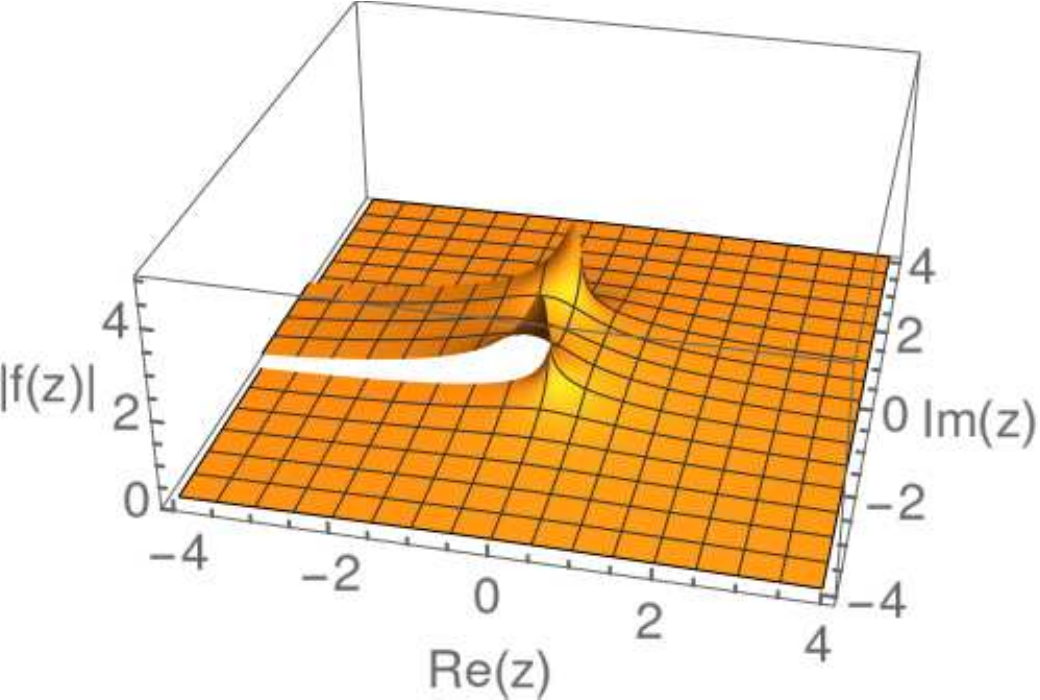}
		\includegraphics[width=5cm]{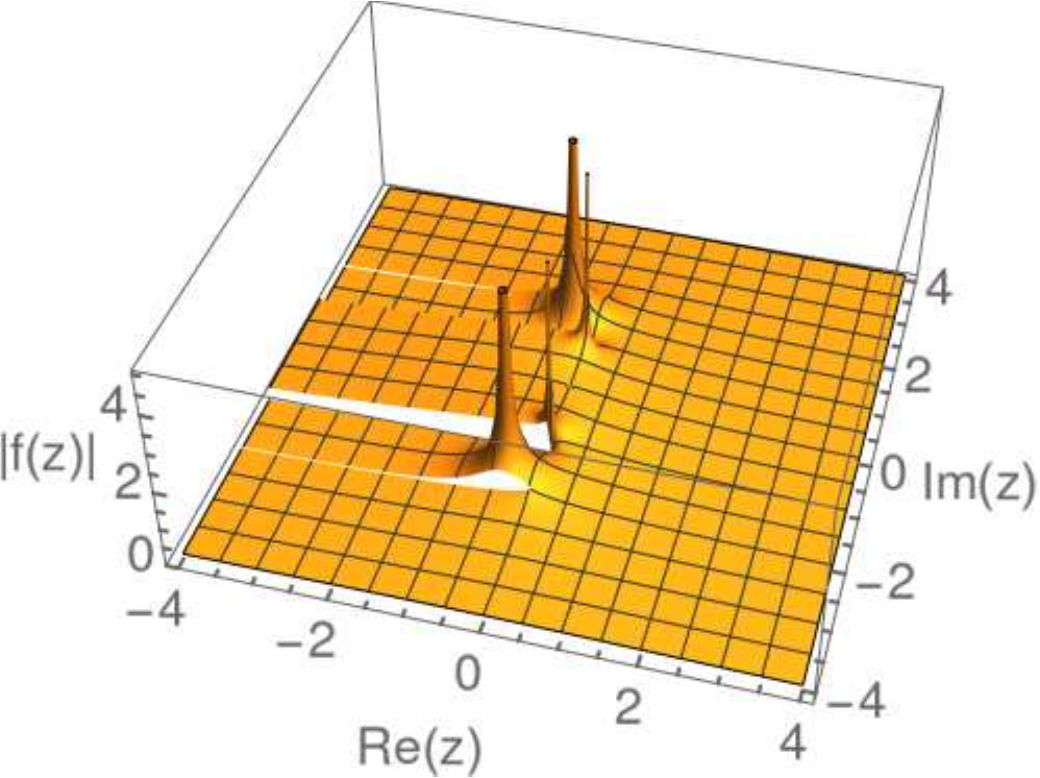}
		\includegraphics[width=5cm]{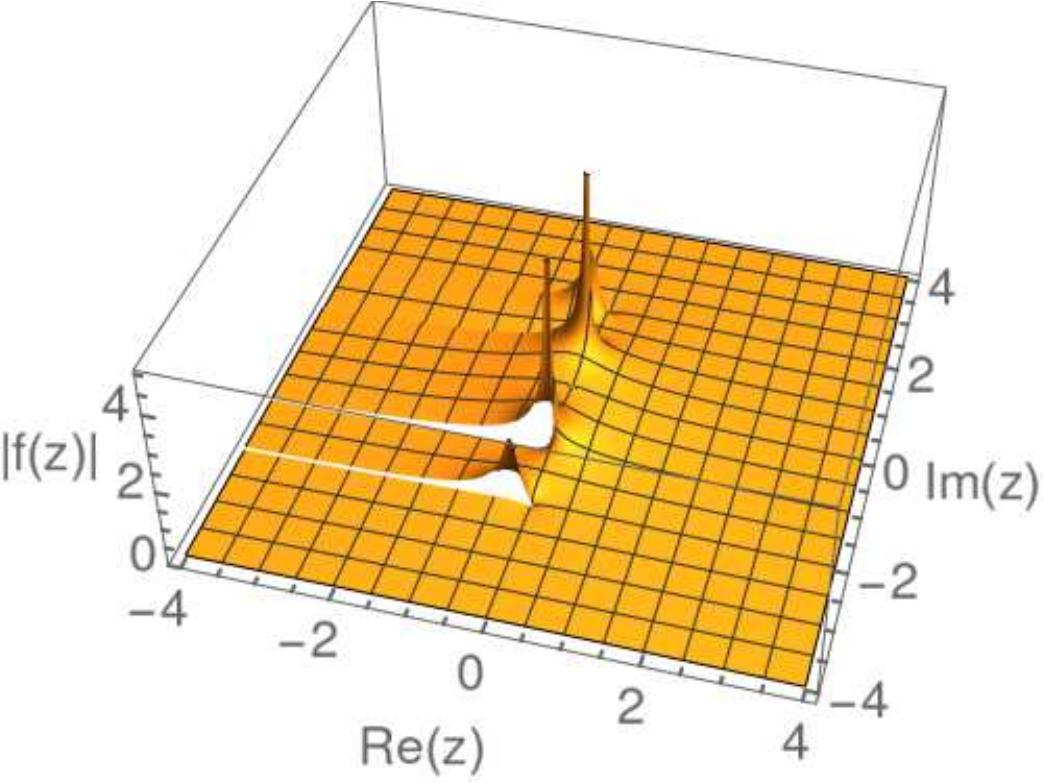}
		\caption{\label{S_fig:fig5} Singularities of the function $\mathcal{R}_{\rm a}(s)$, denoted here as $f(z)$, for $\gamma=2$, and $\alpha=0.5$ (for better display of the cuts). From left to right: screw, glide edge, climb edge. The four poles of the `edge' cases lie close to the branch cuts.}
	\end{minipage}
\end{figure}

To this aim a particular determination must be attributed to square roots in expressions of $\mathcal{C}_{\text{a}}(s)$. Consider for instance the screw-dislocation case where, from Eqs.\ (2), (A4c), and (B4) of the main text 	 
\begin{align}
\label{eq:screwR}
\mathcal{R}_\text{s}(s)&=\frac{1}{\alpha s+\sqrt{1+s^2}}.
\end{align}
The function $\smash{\sqrt{1+s^2}=\sqrt{(s+\ii)(s-\ii)}}$ has branch points $s=\pm\ii$. We fix its determination by taking the branch cuts of the square root on the semi-infinite lines $s=\pm i+\lambda$, with $\lambda<0$. The determination of the argument is taken such that $s-\ii=\rho_1 \ee^{\ii\theta_1}$ and $s+\ii=\rho_2 \ee^{\ii\theta_2}$, with the angles $\theta_1$, $\theta_2$ in $]-\pi,\pi[$. Since $\smash{\sqrt{(s+\ii)(s-\ii)}=\sqrt{1+s^2}=\sqrt{\rho_1\rho_2}\ee^{\ii(\theta_1+\theta_2)/2}}$, this choice of branch cuts amounts to computing $\smash{\sqrt{1+s^2}}$ as $\sqrt{s+\ii}\sqrt{s-\ii}$, using the principal determination of the square root in the last expression [s4]. For the `edge' cases [Eqs.\ (A4f) and (A4i) of the main text] additional branch points are $s=\pm\ii\gamma$, and one would similarly take $\smash{\sqrt{1+s^2/\gamma^2}\to \sqrt{s/\gamma+\ii}\sqrt{s/\gamma-\ii}}$ (recall that $\gamma=\cL/\cS$). Thus, each wavespeed generates two cuts of opposite imaginary part parallel to the negative real axis. Plots of $|\mathcal{R}_{\rm a}(z)|$ using expressions of Appendix A and square roots interpreted thus are displayed in Fig.\ \ref{S_fig:fig5}.

\begin{figure}[!ht]
	\centering
	\begin{minipage}[c]{16cm}
		\includegraphics[width=15cm]{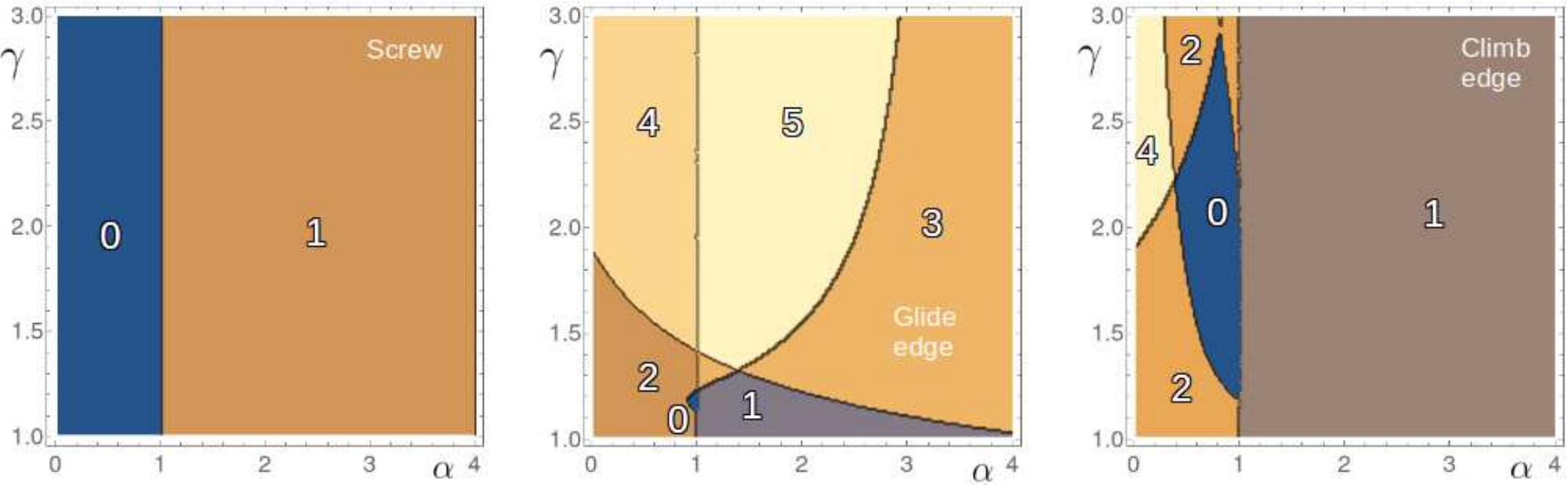}
		\caption{\label{S_fig:fig6} Number of poles of $\mathcal{R}_{\text{a}}(z)$ in parameter space $(\alpha,\gamma)$.}
	\end{minipage}
\end{figure}

Turning now to the poles $\mathcal{R}_{\text{a}}(s)$, their numbers in $(\alpha,\gamma)$ parameter space are reported in Fig.\ \ref{S_fig:fig6} for completeness (recall however that the physical order of magnitude of $\alpha$ is $\alpha\simeq 0.01$, and that $\gamma$ is of order 2). For all three dislocation characters, the value $\alpha=1$ is a special one, if $\alpha$ is introduced as in Eq.\ (2) of the main text. In region $\alpha<1$, all poles occur in conjugate pairs of nonzero imaginary part, wheras in region $\alpha>1$ one of the poles is real negative, which indicates overdamped relaxation. All poles are simple ones, of non-positive real part. 

Kernel $\mathcal{R}_{\text{s}}(s)$ has only one pole, namely $s=-1/\sqrt{\alpha^2-1}$ if $\alpha>1$, and no pole otherwise. In the 'edge' cases the algebraic analytical expressions of $\mathcal{R}_{\text{c,g}}(s)$ allow for a determination of the poles by solving numerically a 12th-degree polynomial equation, and by selecting the relevant solutions by susbtitution into the denominator of $\mathcal{R}_{\text{c,g}}(s)$. All poles are located in-between the cuts $s\in]-\infty\pm\ii\gamma,\pm\ii\gamma]$.
\bigskip

{\bf References:}
\medskip

[s1] L.~Pillon, C.~Denoual, and Y.-P.~Pellegrini,
Equation of motion for dislocations with inertial effects, \href{http://prb.aps.org/abstract/PRB/v76/i22/e224105}{Phys. Rev. B \textbf{76}, 224105
	(2007)}.

[s2] B.~Gurrutxaga--Lerma, J.~Verschueren, A.P.~Sutton, and D.\ Dini,
The mechanics and physics of high-speed dislocations: a critical review,
\href{https://doi.org/10.1080/09506608.2020.1749781}{Int.\ Mater.\ Rev.\ \textbf{66},
	215--255 (2020)}

[s3] J.~P.~Hirth, H.~M.~Zbib, and J.~Lothe,
Forces on high-velocity dislocations,
\href{https://doi.org/10.1088/0965-0393/6/2/006}{Model.\ Simul.\ Mat.\ Sci.\ Eng.\ \textbf{6}, 165 (1998).}

[s4] D.~G.~Duffy,
\href{https://doi.org/10.1201/9781420035148}{\emph{Transform Methods for Solving Partial Differential Equations, 2nd ed.}},
(Chapman and Hall/CRC, New York, 2004).
\end{document}